\documentclass[aps,twocolumn,showpacs,superscriptaddress]{revtex4-1}
\usepackage[english]{babel}
\usepackage{graphicx,dcolumn,bm,amssymb,amsmath,amsfonts,amsthm,latexsym}
\usepackage{epstopdf}
\usepackage{float,subfig,slashed,hyperref}
\usepackage[toc]{appendix}
\hypersetup{
  linktocpage  = true,
  colorlinks   = true,
  urlcolor     = red,
  linkcolor    = black,
  citecolor    = blue
}
\begin{document}
\widetext

\title{Vector and axial-vector meson properties\\ in a nonlocal SU(2) PNJL model}
\author{J.P.~Carlomagno}
\email{carlomagno@fisica.unlp.edu.ar}
\affiliation{IFLP, CONICET $-$ Departamento de F\'{i}sica, Universidad Nacional de La Plata, C.C. 67, 1900 La Plata, Argentina}
\affiliation{CONICET, Rivadavia 1917, 1033 Buenos Aires, Argentina}
\author{M.F.~Izzo~Villafa\~ne}
\email{izzo@tandar.cnea.gov.ar}
\affiliation{CONICET, Rivadavia 1917, 1033 Buenos Aires, Argentina}
\affiliation{Physics Department, Comisi\'on Nacional de Energ\'ia At\'omica, Av. Libertador 8250, (1429) Buenos Aires, Argentina}

\begin{abstract}
We study the features of a SU(2) Polyakov-Nambu-Jona-Lasinio model that includes wave function renormalization and nonlocal vector interactions.
Within this framework we analyze, among other properties, the masses, width and decay constants of light vector and axial-vector mesons at finite temperature.
Then we obtain the corresponding phase diagram in a finite density scenario, after characterizing the deconfinement and chiral restoration transitions.
\end{abstract}

\maketitle

\section{Introduction}
\label{intro}

The phase diagram of strongly interacting matter at finite temperature $T$ and chemical potential $\mu$ has been extensively studied along the past decades. 
Quantum Chromodynamics (QCD) predicts that at very high temperatures ($T \gg \Lambda_{\rm QCD}$) and low densities this matter appears in the form of a plasma of quarks and gluons~\cite{Fukushima:2010bq}. 
At such extreme conditions, QCD is weakly coupled and first-principle perturbative calculations based on an expansion in the coupling constant can be used to explore the phase diagram. 
However, in the low-energy regime the analysis of hadron phenomenology starting from first principles is still a challenge for theoretical physics. 
Although substantial progress has been achieved in this sense through lattice QCD (lQCD) calculations, this approach shows significant difficulties, e.g.~when dealing with small current quark masses and/or finite chemical potential. 
Thus, most of the present knowledge about the behavior of strongly interacting matter arises from the study of effective models, which offer the possibility to get predictions of the transition features at regions that are not accessible through lattice techniques.

Here we will concentrate on one particular class of effective theories, viz. the nonlocal Polyakov$-$Nambu$-$Jona-Lasinio (nlPNJL) models (see Refs.~\cite{Villafane:2016ukb, Carlomagno:2018tyk} and references therein), in which quarks interact through covariant nonlocal chirally symmetric four point couplings in a background color field, and gluon self-interactions are effectively introduced by a Polyakov loop effective potential.
These approaches, which can be considered as an improvement over the (local) PNJL model, offer a common framework to study both the chiral restoration and deconfinement transitions. 
In fact, the nonlocal character of the interactions leads to a momentum dependence in
the quark propagator that can be made consistent~\cite{Noguera:2008} with
lattice results.

This scheme has been used to describe the chiral restoration transition for hadronic systems at finite temperature and/or chemical potential (see e.g.\ Refs.~\cite{GomezDumm:2001fz,Hell:2008cc,Radzhabov:2010dd,Contrera:2010kz,Hell:2011ic,Carlomagno:2013ona,Carlomagno:2018tyk}).
In this work, following Ref.~\cite{Villafane:2016ukb}, we concentrate in the incorporation of vector and axial-vector interactions extended to finite $T$ and $\mu$. 
Therefore, besides the scalar and pseudoscalar quark-antiquark currents, we include couplings between vector and axial-vector nonlocal currents satisfying proper QCD symmetry requirements.

Within this theoretical framework, we study the thermal behavior of several properties for the vector meson $\rho$ and the axial-vector meson $\rm a_1$.
We start by analyzing the temperature dependence of their masses, decay constants and decay widths, and then we characterize the chiral and deconfinement transitions through the corresponding order parameters to finally obtain the QCD phase diagram at finite chemical potential.

This article is organized as follows. 
In Sect.~\ref{model} we present the general formalism to describe a system at finite temperature and density. 
The numerical and phenomenological analyses for several meson properties and the corresponding QCD phase diagram are included in Sect.~\ref{results} and~\ref{QCDpd}. 
Finally, in Sect.~\ref{summary} we summarize our results and present the conclusions. 

\section{Thermodynamics}
\label{model}

We consider a two-flavor chiral quark model that includes nonlocal vector and axial-vector quark-antiquark currents. 
The corresponding Euclidean effective action and nonlocal fermion currents can be found in Ref.~\cite{Villafane:2016ukb}.

We perform a bosonization of the fermionic theory~\cite{Ripka:1997zb} in a standard way by considering the partition function
$\mathcal{Z} = \int \mathcal{D}\, \bar{\psi}\mathcal{D}\psi \,\exp[-S_E]$
and introducing auxiliary fields. 
After integrating out the fermion fields the partition function can be written in the mean field approximation (MFA), in which the bosonic fields are expanded around their vacuum expectation values, $\phi(x) = \bar \phi + \delta\phi(x)$. 

We extend now the analysis of this model to a system at finite temperature and chemical potential.
Following the same prescriptions described in Refs.~\cite{GomezDumm:2001fz,GomezDumm:2004sr} the thermodynamic potential in the MFA is given by~\cite{Contrera:2012wj}
\begin{equation}
\label{omegareg}
\Omega^{\rm MFA} \ = \ \Omega^{\rm reg} + \Omega^{\rm free} +
\mathcal{U}(\Phi,T) + \Omega_0 \ ,
\end{equation}
where
\begin{widetext}
\begin{align}
\Omega^{\rm reg} &=  \,- \,4 T \sum_{c=r,g,b} \ \sum_{n=-\infty}^{\infty}
\int \frac{d^3\vec p}{(2\pi)^3} \ \ln \left[ \frac{ (\rho_{n,
\vec{p}}^c)^2 + m^2(\rho_{n,\vec{p}}^c)}{z^2(\rho_{n, \vec{p}}^c)\ [(\rho_{n,
\vec{p}}^c)^2 + m^2]}\right]+
\frac{\bar\sigma_1^2 + \kappa_p^2\; \bar\sigma_2^2}{2\,G_S} - \frac{\bar\omega^2}{2\,G_0} \ , \nonumber \\
\Omega^{\rm free} \ &= \ -4 T \sum_{c=r,g,b} \ \sum_{s=\pm 1} \int \frac{d^3 \vec{p}}{(2\pi)^3}\; \mbox{Re}\;
\ln \left[ 1 + \exp\left(-\;\frac{\epsilon_p + s(\mu + i \phi_c)}{T}
\right)\right]
\ .
\label{granp}
\end{align}
\end{widetext}
The constants $\bar\sigma_{1,2}$ and $\bar\omega$ are the mean field values of the scalar and the isospin zero vector fields.
At nonzero quark densities, the flavor singlet term of the vector interaction develops a nonzero expectation value, while all other components of the vector and axial vector interactions have vanishing mean fields~\cite{Bratovic:2012qs}.

The mean field values can be calculated by minimizing $\Omega^{\rm MFA}$, while $m(p)$ and $z(p)$ are the momentum-dependent effective mass and the wave function renormalization (WFR). 
These functions are related to the nonlocal form factors and the vacuum expectation values of the scalar fields by
\begin{eqnarray}
m(p) &=& z(p)\, \left[ m\, +\, \bar \sigma_1\, g(p)\right] \ ,\nonumber\\
z(p) &=& \left[ 1\,-\,\bar \sigma_2 \,f(p)\right]^{-1}\ ,
\label{mz}
\end{eqnarray}
where $g(p)$ and $f(p)$ are the Fourier transforms of the form factors included in the nonlocal quark antiquark currents (see Ref.~\cite{Villafane:2016ukb}).
We have also defined
\begin{equation}
\Big({\rho_{n,\vec{p}}^c} \Big)^2 =
\Big[ (2 n +1 )\pi  T + \phi_c - \imath \tilde{\mu} \Big]^2 + {\vec{p}}\ \! ^2 \ , 
\end{equation}
in which the sums over color indices run over $c=r,g,b$ with the color background fields components being $\phi_r = - \phi_g = \phi$, $\phi_b = 0$, and $\epsilon_p = \sqrt{\vec{p}^{\;2}+m^2}\;$. 
The vector coupling generates a shifting in the chemical potential as~\cite{Contrera:2012wj}
\begin{equation}
\tilde{\mu} = \mu - g({\rho_0}_{n,\vec{p}}^c)\ z({\rho_0}_{n,\vec{p}}^c)\ \bar{\omega} \ ,
\end{equation}
where 
\begin{equation}
{\rho_0}_{n,\vec{p}}^c = \rho_{n,\vec{p}}^c\ \vert_{\bar{\omega}=0} \ .
\end{equation}

The term $\Omega^{\rm reg}$ is obtained after following the same regularization prescription as in previous works~\cite{GomezDumm:2001fz}. 
Namely, we subtract the thermodynamic potential of a free fermion gas, and then we add it in a regularized form. 
Finally, the last term in Eq.~(\ref{omegareg}) is a constant fixed by the condition that $\Omega^{\rm MFA}$ vanishes at $T=\mu=0$.

The chiral quark condensates are given by the vacuum expectation values $\langle \bar qq \rangle$. 
The corresponding expressions can be obtained by differentiating $\Omega^{\rm MFA}$ with respect to the current quark masses.

In this effective model we assume that fermions move on a static and constant background gauge field $\phi$, the Polyakov field, that couples with fermions through the covariant derivative in the fermion kinetic term.
The traced Polyakov loop (PL), $\Phi=\frac{1}{3} {\rm Tr}\, \exp( i \phi/T)$, can be considered as the order parameter for confinement in the infinite quark mass limit ~\cite{tHooft:1977nqb, Polyakov:1978vu}.
The effective gauge field self-interactions are given by the Polyakov loop potential $\mathcal{U}(\Phi,T)$, whose functional form is usually based on properties of pure gauge QCD. 
Among the most used effective potentials, the Ansatz that provides a good agreement with lQCD results~\cite{Carlomagno:2013ona,Carlomagno:2018tyk} is a polynomial function based on a Ginzburg-Landau Ansatz~\cite{Ratti:2005jh,Scavenius:2002ru}:
\begin{align}
\frac{{\cal{U}}_{\rm poly}(\Phi ,T)}{T ^4} \ = \ -\,\frac{b_2(T)}{2}\, \Phi^2
-\,\frac{b_3}{3}\, \Phi^3 +\,\frac{b_4}{4}\, \Phi^4 \ ,
\label{upoly}
\end{align}
where
\begin{align}
b_2(T) = a_0 +a_1 \left(\dfrac{T_0}{T}\right) + a_2\left(\dfrac{T_0}{T}\right)^2
+ a_3\left(\dfrac{T_0}{T}\right)^3\ .
\label{pol}
\end{align}
The numerical values for the parameters can be found in Ref.~\cite{Ratti:2005jh}.

From lattice calculations one would expect to find a deconfinement temperature of $T_0 = 270$~MeV given the absence of dynamical quarks.
However, it has been argued that in the presence of light dynamical quarks this temperature scale, which is a further parameter of the model, should be adequately reduced to about $210$ and $190$~MeV for the case of two and three flavors respectively, with an uncertainty of approximately $30$~MeV~\cite{Schaefer:2007pw}.

\subsection{Meson Masses}
In general, meson masses can be obtained from the terms in the Euclidean action that are quadratic in the bosonic fields.
The resulting scalar and pseudoscalar meson sector of the bosonized Euclidean action can be writen as~\cite{Noguera:2008}
\begin{eqnarray} 
  S^{\rm S,PS}_E = \dfrac{1}{2} \int \frac{d^4 p}{(2\pi)^4}\
      G_K(p^2)\, \delta K(p)\, \delta K(-p) 
\end{eqnarray}
with $K= \sigma, \sigma'$ and $\vec \pi$.

In the vector meson sector we find instead~\cite{Villafane:2016ukb}
\begin{eqnarray} \label{eq:quad}
S_E^{\rm V,A} &=& \dfrac{1}{2} \int \frac{d^4 p}{(2\pi)^4}\
\Big\{ G_V^{\mu\nu}(p^2)\,\delta V_\mu(p)\cdot\delta V_\nu(-p) \nonumber\\
&& \hspace{-1.2cm} + \,
i\,G_{\pi a}(p^{2})\Big[p^\mu\,\delta\vec{a}_{\mu}(-p) \cdot
                \delta\vec{\pi}(p)-p^\mu\,\delta\vec{a}_{\mu}(p)
                \cdot \delta\vec{\pi}(-p)\Big]                
\Big\} , \nonumber\\
\end{eqnarray}
with $V_\mu = v_\mu^0, a_\mu^0, \vec v_\mu$, and $\vec a_\mu$. 
For this case we obtain the tensors $G_{v^0}^{\mu\nu}$, $G_{a^0}^{\mu\nu}$, $G_v^{\mu\nu}$ and $G_a^{\mu\nu}$ from the expansion of the fermionic determinant. 
One has
\begin{eqnarray}
G_{v,a}^{\mu\nu}(p^2) &=& G_{\rho,{\rm a}_1}(p^2)\left(\delta^{\mu\nu}-\dfrac{p^{\mu}p^{\nu}}{p^2}\right)+ L_\pm (p^2)\dfrac{p^{\mu}p^{\nu}}{p^{2}} ,\nonumber \\
\end{eqnarray}
where the functions $G_{\rho,{\rm a}_1}(p^2)$ and $L_\pm (p^2)$ correspond to the transverse and longitudinal projections of the vector and axial-vector fields, describing meson states with spin 1 and 0, respectively.
Regarding the isospin zero channels, it is easy to see that the expressions for $G_{v^0}^{\mu\nu}$ and $G_{a^0}^{\mu\nu}$ can be obtained from $G_v^{\mu\nu}$ and $G_a^{\mu\nu}$ simply by replacing the vector coupling constant $G_V\rightarrow G_0$ and $G_V\rightarrow G_5$ in each case.
It can also be seen in Eq.~(\ref{eq:quad}) that due to the addition of the vector meson sector, a mixing between the pion fields and the longitudinal part of the axial-vector fields arises~\cite{Ebert:1985kz,Bernard:1993rz}.
To remove this mixing and obtain the correct pion mass of the physical state $\tilde{\vec \pi}$ we introduce a mixing function $\lambda(p^2)$, defined in such a way that the cross terms in the quadratic expansion vanish~\cite{Villafane:2016ukb}.

Once cross terms have been eliminated, the resulting functions $G_M(p^2)$ stand for the inverses of the effective meson propagators.

At finite temperature, the meson masses are calculated by solving $G_M (p_M^2) = 0$ with $p_M = (0,im_M)$ (see Appendix~\ref{App_masses_decays} for the analytical expressions). 
The mass values determined by these equations are the spatial ``screening-masses'' corresponding to the zeroth Matsubara mode, and their inverses describe the persistence lengths of these modes at equilibrium with the heat bath~\cite{Contrera:2009hk}.

Finally, the meson wave function renormalization and the meson-quark effective coupling constant can be obtained from
\begin{equation}
\label{zpi} 
Z_M^{-1} \, = \, g_{Mqq}^{-2} \,= \,  \frac{dG_M(p^2)}{dp^2}\bigg\vert_{p^2=-m_M^2} \ ,
\end{equation}
and thus the physical states of the meson fields can be calculated.

\subsection{Decay constants}

The pion weak decay constant $f_\pi$ is given by the matrix elements of axial currents between the vacuum and the physical one-pion states at the pion pole,
\begin{equation}
\label{eq:fpi} 
\langle 0 \vert J_{A\mu}^a (0) \vert \tilde{\pi}^b(p) \rangle = 
i \, \delta^{ab} \, f_\pi(p^2) \; p_\mu \ .
\end{equation}

On the other hand, the matrix elements of the electromagnetic current $J_{em}$ between the neutral vector meson state and the vacuum determine the vector decay constant $f_v$
\begin{equation}
\label{eq:frho} 
\langle 0 \vert J_{em\ \mu} (0) \vert \rho_\nu^0 (p) \rangle = 
e \, f_v(p^2) (\delta_{\mu\nu} p^2 - p_\mu p_\nu) \ ,
\end{equation}
where $e$ is the electron charge. 
We can easily notice that, as required from the conservation of the electromagnetic current, the matrix element is transverse.

In order to obtain these matrix elements within our model, we have to gauge the effective action through the introduction of gauge fields, and then calculate the functional derivatives of the bosonized action with respect to the currents and the renormalized meson fields.
In addition, due to the nonlocality of the interaction, the gauging procedure requires the introduction of gauge fields not only through the usual covariant derivative in the Euclidean action but also through a transport function that comes with the fermion fields in the nonlocal currents (see e.g. Refs.~\cite{Ripka:1997zb,Bowler:1994ir,GomezDumm:2006vz}).

After a lengthy calculation described in Ref~\cite{Villafane:2016ukb}, we find that the decay constants at $T=0$ are given by
\begin{eqnarray}
f_\pi &=& \dfrac{m_q \, Z_\pi^{1/2}}{m_\pi^2} \left[ F_0 (-m_\pi^2)
+\dfrac{G_{\pi a}(p^2)}{L_-(p^2)} \, F_1 (-m_\pi^2)\right] \ ,\nonumber\\
f_v \ &=& \ \dfrac{Z_\rho^{1/2}}{3\, m_\rho^2}
\, \left[ J_V^{\rm (I)} (-m_\rho^2) + J_V^{\rm (II)} (-m_\rho^2)\right]\ .
\label{efes}
\end{eqnarray}
The detailed analytical expressions for the functions $F_0$, $F_1$ and $J_{V}^{\rm (I, II)}$ at finite temperature can be found in Appendix~\ref{App_masses_decays}.

Moreover, another important quantity that can be studied within the effective model is the axial-vector decay constant $f_{\rm a}$, which is defined from the matrix elements of the electroweak charged currents $J_{ew}$ between the axial-vector meson state and the vacuum at $p^2 = -m_{\rm a_1}^2$. 
We have
\begin{equation}
\langle 0 \vert J_{ew\ \mu} (0) \vert a_{1 \nu}(p)\rangle  = \Pi_{\mu\nu}(p^2) \ ,
\end{equation}
with $\Pi_{\mu\nu}  =  T(p^2) \big( \delta_{\mu\nu}\,p^2 \, - \, p_\mu\,p_\nu \big)\  +\ L(p^2) \, p_\mu\, p_\nu$.
Given the mixing between the pion field and the axial-vector field, the longitudinal projection of $\Pi_{\mu\nu}$ contributes to the pion weak decay constant. 
The term from which $f_{\rm a}$ can be calculated is the transverse projection, which corresponds to the physical state $\tilde{\vec{a}}_\mu$. 
Therefore, we define $f_{\rm a} = T(p^2)\vert_{p^2=-m_{{\rm a}_1}^2}$ and find the following expression
\begin{eqnarray}
\label{eq:fa}
 f_{\rm a} \ &=& \ \dfrac{Z_{{\rm a}_1}^{1/2}}{3\, m_{{\rm a}_1}^2}
\, \left[ J_A^{\rm (I)} (-m_{{\rm a}_1}^2) + J_A^{\rm (II)} (-m_{{\rm a}_1}^2)\right]\ ,
\end{eqnarray}
where

\begin{widetext}
\begin{eqnarray}
J_A^{\rm (I)}  (p^2) &=&  -\,4N_c\,\int \dfrac{d^4 q}{(2\pi)^4}\, g(q)\,\Bigg\lbrace
              \dfrac{3}{2}\,\dfrac{[z(q^+)+z(q^-)]}{D(q^+)D(q^-)}\Big[
              (q^+\cdot q^-) - m(q^+)\,m(q^-) \Big] \nonumber \\
              & & + \; \dfrac{1}{2}\,\dfrac{z(q^+)}{D(q^+)} \, +
                  \, \dfrac{1}{2}\, \dfrac{z(q^-)}{D(q^-)}\, - \,\dfrac{q^2}{(q\cdot p)}
                  \left[\dfrac{z(q^+)}{D(q^+)} - \dfrac{z(q^-)}{D(q^-)}\right]\nonumber \\
              & & + \,\dfrac{z(q^+)z(q^-)}{D(q^+)D(q^-)}\,
              \left[(q\cdot p) - \dfrac{q^2\,p^2}{(q\cdot p)}\right]\,
                  \bigg[\bar\sigma_1\, \big[m(q^-) - m(q^+)\big]\,\alpha^+_g(q,p) \nonumber \\
              & & + \; \bar\sigma_2\,\big[q^2 + \dfrac{p^2}{4}
                  + m(q^+)\,m(q^-) \big]\,\alpha^-_f (q,p)\,  + 
                  \frac{2}{q^2}\left(\bar\sigma_1\ g(q) + m_q \right)\big[m(q^-) - m(q^+)\big]\ \bigg] \Bigg\rbrace\ , \nonumber\\  
J_A^{\rm (II)}  (p^2) &=&   -\,4N_c \int \dfrac{d^4 q}{(2\pi)^4}\,
              \dfrac{z(q)}{D(q)}\left\lbrace\dfrac{q^2}{(q\cdot p)}
              \Big[g(q^+)-g(q^-)\Big] + \left[(q\cdot p) -
              \dfrac{q^2\,p^2}{(q\cdot p)}\right]\alpha^-_g (q,p)
              \right\rbrace ,
\end{eqnarray}
\end{widetext}

and 
\begin{equation}
  \alpha_f^\pm(q,p)  =  \int\limits^1_0 d\lambda\, \dfrac{\lambda}{2}\, \left[f'\left(q+\lambda\dfrac{p}{2}\right)\ \pm \ 
		  f'\left(q-\lambda\dfrac{p}{2}\right)\right].
\end{equation}

\subsection{Decay widths}

Various transition amplitudes can be calculated by expanding the bosonized action to higher orders in meson fluctuations. 
In this work we are particularly focused in the study of the processes $\rho\to\pi\pi$~\cite{Villafane:2016ukb} and ${\rm a}_1\to\rho\pi$ at finite temperature.
The decay amplitudes $\mathcal{A}_\rho(v^a(p)\to \pi^b(q_1)\pi^c(q_2))$ and $\mathcal{A}_{{\rm a}_1}(a^a(p)\to v^b(q_1)\pi^c(q_2))$ are obtained by  calculating the corresponding functional derivatives of the effective action.
Thus, we define
\begin{eqnarray}
	\frac{\delta^3 S_E^{\rm bos}}{\delta\tilde v_\mu^a(p) \delta
	 \tilde{\pi}^b(q_1) \delta\tilde{\pi}^c(q_2)} \bigg\vert_{\delta v_\mu=\delta\pi=0} 
		&=& \nonumber\\
		& & \hspace{-2.5cm} \hat \delta^{(4)} (p + q_1 + q_2) \ \epsilon_{abc} 
		    \; V^{\rho \rightarrow \pi\pi}_\mu \nonumber\\
	\frac{\delta^3 S_E^{\rm bos}}{\delta\tilde a_\mu^a(p) \delta\tilde{v}_\nu^b(q_1)
	 \delta\tilde{\pi}^c(q_2) } \bigg\vert_{\delta a_\mu=\delta v_\nu=\delta\pi=0} 
		&=& \nonumber\\
		& & \hspace{-2.5cm} \hat \delta^{(4)} (p + q_1 + q_2) \ \epsilon_{abc} 
		    \; \Pi^{\rm a_1 \rightarrow \rho\pi}_{\mu\nu}, \nonumber\\		
\end{eqnarray}
where $\hat \delta^{(4)}(p + q_1 + q_2) = (2\pi)^4\,\delta^{(4)}(p + q_1 + q_2)$, and
\begin{eqnarray}
    \label{eq:matrix}
    V^{\rho \rightarrow \pi\pi}_\mu &=& \tilde{F}_{\rho\pi\pi}(p^2,q_1^2,q_2^2)\dfrac{(q_{1\mu} + q_{2\mu})}{2} \nonumber\\
				    & & + \,
					\tilde{G}_{\rho\pi\pi}(p^2,q_1^2,q_2^2)\dfrac{(q_{1\mu} - q_{2\nu})}{2} \ , \nonumber\\
    \Pi^{\rm a_1 \rightarrow \rho\pi}_{\mu\nu} &=& \tilde{F}_1(p^2,q_1^2,q_2^2)\, \delta_{\mu\nu} \, + \, 
						    \tilde{F}_2(p^2,q_1^2,q_2^2)\, q_{1\mu} q_{1\nu} \nonumber\\
					& &	  + \, \tilde{F}_3(p^2,q_1^2,q_2^2)\, q_{2\mu} q_{2\nu}
		      \, +\, \tilde{F}_4(p^2,q_1^2,q_2^2)\, q_{1\mu} q_{2\nu} \nonumber\\
		      & & + \, \tilde{F}_5(p^2,q_1^2,q_2^2)\, q_{2\mu} q_{1\nu} .
\end{eqnarray}
The form factors $\tilde G_{\rho\pi\pi}(p^2,q_1^2, q_2^2)$, $\tilde F_{\rho\pi\pi}(p^2,q_1^2, q_2^2)$ and $\tilde F_i(p^2, q_1^2, q_2^2)$ are one-loop functions that arise from the expansion of the effective action, and the explicit forms of those used for our calculations can be found in Appendix~\ref{App_widths}. 

Finally, after some algebra we obtain the following expressions for the amplitudes 
\begin{eqnarray}
\label{eq:amplitude}
\vert \mathcal{A}_\rho \vert^2 &=& \dfrac{m_\rho^2}{3} \bigg(1-\dfrac{4m_\pi^2}{m_\rho^2}\bigg) g_{\rho\pi\pi}^2  \ , \nonumber\\
\vert \mathcal{A}_{{\rm a}_1} \vert^2 &=& 2\,g_{{\rm a}\rho\pi}^2 + \dfrac{1}{16\,m_\rho^2m_{{\rm a}_1}^2} \Big\{2\,g_{{\rm a}\rho\pi}(m_{{\rm a}_1}^2-m_\pi^2+m_\rho^2) \nonumber\\
				& & \hspace{-0.5cm} +
	  f_{{\rm a}\rho\pi}\big[m_{{\rm a}_1}^4-2\,m_{{\rm a}_1}^2(m_\rho^2+m_\pi^2)+(m_\rho^2-m_\pi^2)^2\big]\Big\}^2\ ,\nonumber\\
\end{eqnarray}
with $g_{\rho\pi\pi}\equiv\tilde G_{\rho\pi\pi}(-m_\rho^2,-m_\pi^2,-m_\pi^2)$,
$g_{{\rm a}\rho\pi}\equiv\tilde F_1(-m_{{\rm a}_1}^2,-m_\rho^2,-m_\pi^2)$ and
$f_{{\rm a}\rho\pi}\equiv\tilde F_3(-m_{{\rm a}_1}^2,-m_\rho^2,-m_\pi^2)-\tilde F_4(-m_{{\rm a}_1}^2,-m_\rho^2,-m_\pi^2)$.
It is quite straightforward to see that only the transverse piece of $ V^{\rho \rightarrow \pi\pi}_\mu$ contributes to the $\rho\to\pi\pi$ decay width. 
That is not, however, the case for $\rm a_1\to\rho\pi$.

In order to study the thermal dependence of these decay widths, it is necessary to modify the two-body phase space to include finite temperature effects.
Following Refs.~\cite{Weldon:1991ei,Tanabashi:2018oca}, the decay of a particle at rest of mass $M$, into particles of masses $m_1$ and $m_2$ in equilibrium with the heat bath, is given by
\begin{eqnarray}
\label{gammaofT}
&&\Gamma_{M\to m_1 m_2}\ \big\vert_{p=0} = \frac{\vert \mathcal{A}_M \vert^2}{32 \pi M}\ \times \nonumber \\
&& \sqrt{\left(1-\frac{(m_1 + m_2)^2}{M^2}\right)
\left(1-\frac{(m_1 - m_2)^2}{M^2}\right)}\ \times \nonumber \\
&& \frac{\exp{\left[\frac{1}{2T}M\right]}}{\cosh{\left[\frac{1}{2T}M\right]}-\cosh{\left[\frac{1}{2T}\frac{(m_1-m_2)(m_1+m_2)}{M}\right]}} \ ,
\end{eqnarray}
where $\mathcal{A}_M$ is evaluated within the effective model.
Hence, together with the results of Eq.~(\ref{eq:amplitude}) it leads to
\begin{eqnarray}
\Gamma_{\rho\to \pi\pi} &=& \frac{\vert \mathcal{A}_{\rho} \vert^2}{32 \pi \, m_\rho} 
			    \bigg(1-\dfrac{4m_\pi^2}{m_\rho^2}\bigg)^{1/2} 
			    \frac{ \exp{\left[ \frac{1}{2T}\,m_\rho \right]} }{\cosh{ \left[\frac{1}{2T}\, m_\rho \right]} \ -\ 1 } 
			    \ , \nonumber\\
\Gamma_{{\rm a}_1\to \rho\pi} &=& \frac{\vert \mathcal{A}_{{\rm a}_1} \vert^2}{32 \pi\, m_{{\rm a}_1}^3 } 
 \Big[m_{{\rm a}_1}^2 - (m_\rho + m_\pi)^2\Big]^{1/2} \, \times
  \nonumber \\
&&  \frac{\Big[m_{{\rm a}_1}^2 - (m_\rho - m_\pi)^2\Big]^{1/2} \exp{\left[\frac{1}{2T}m_{{\rm a}_1}\right]}}{\cosh{\left[\frac{1}{2T}m_{{\rm a}_1}\right]}-
\cosh{\left[\frac{1}{2T}\frac{(m_\rho-m_\pi)(m_\rho+m_\pi)}{m_{{\rm a}_1}}\right]}} \ . \nonumber\\
\end{eqnarray}

\section{Numerical Results}
\label{results}

\subsection{Vacuum properties} 

To fully define the model it is necessary to specify the form factors $f(z)$ and $g(z)$ in the nonlocal fermion currents. 
In this work we will consider an exponential momentum dependence, which guarantee a fast ultraviolet convergence of the loop integrals,
\begin{eqnarray*}
&& g(p) = {\rm exp}(-p^2 / \Lambda_0^2)\ , \quad
   f(p) = {\rm exp}(-p^2 / \Lambda_1^2)\  .
\label{gyf}
\end{eqnarray*}
Notice that in these form factors two energy scales, $\Lambda_0$ and $\Lambda_1$, are introduced, which act as effective momentum cutoff and have to be taken as additional parameters of the model. 
Moreover, the model in the MFA includes other five free parameters, namely the current quark mass $m$, the coupling constants $G_S$, $G_V$, $G_0$ and $\kappa_p$.

\begin{figure}[t]
\centering
\subfloat{\includegraphics[width=0.4\textwidth]{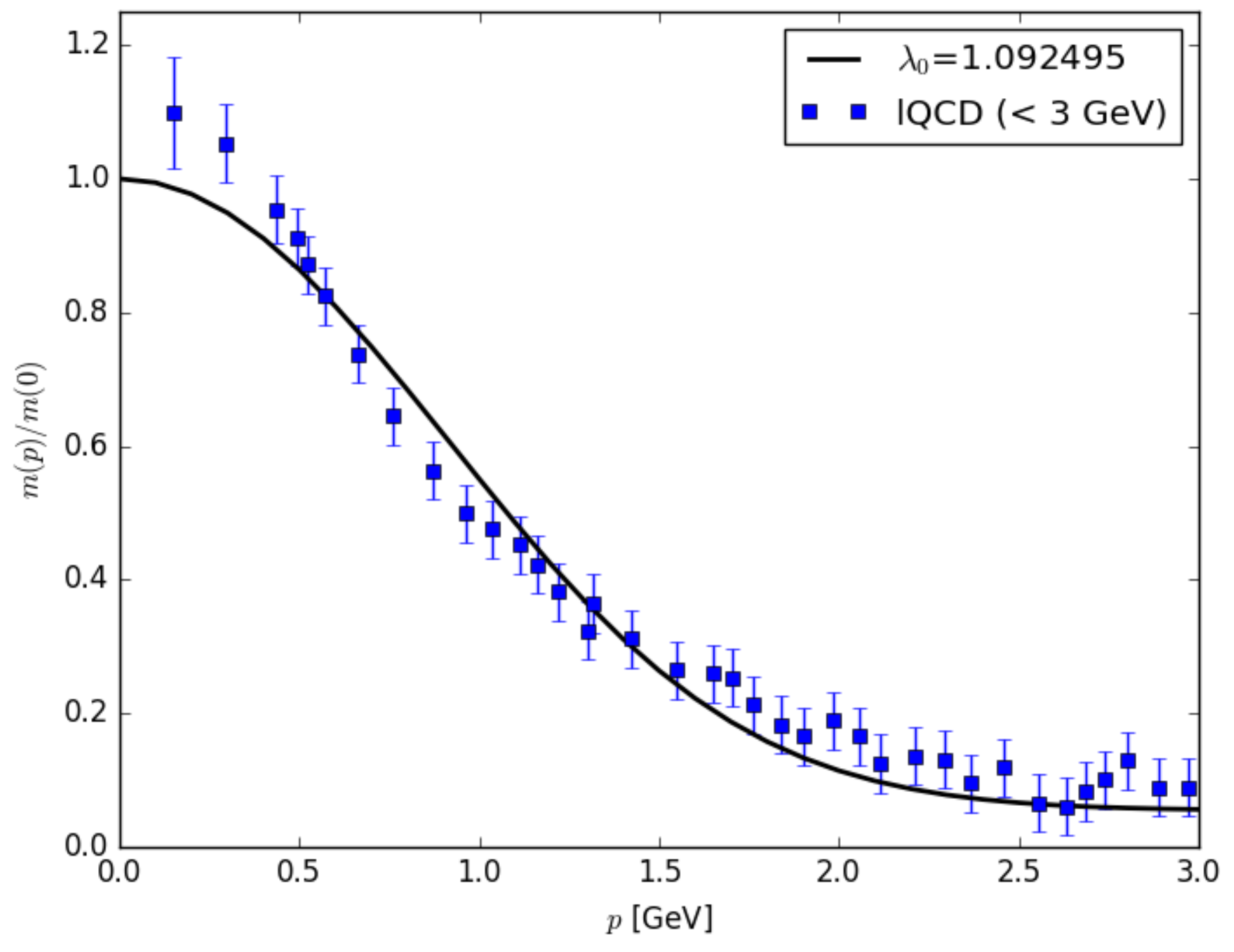}}\\
\subfloat{\includegraphics[width=0.4\textwidth]{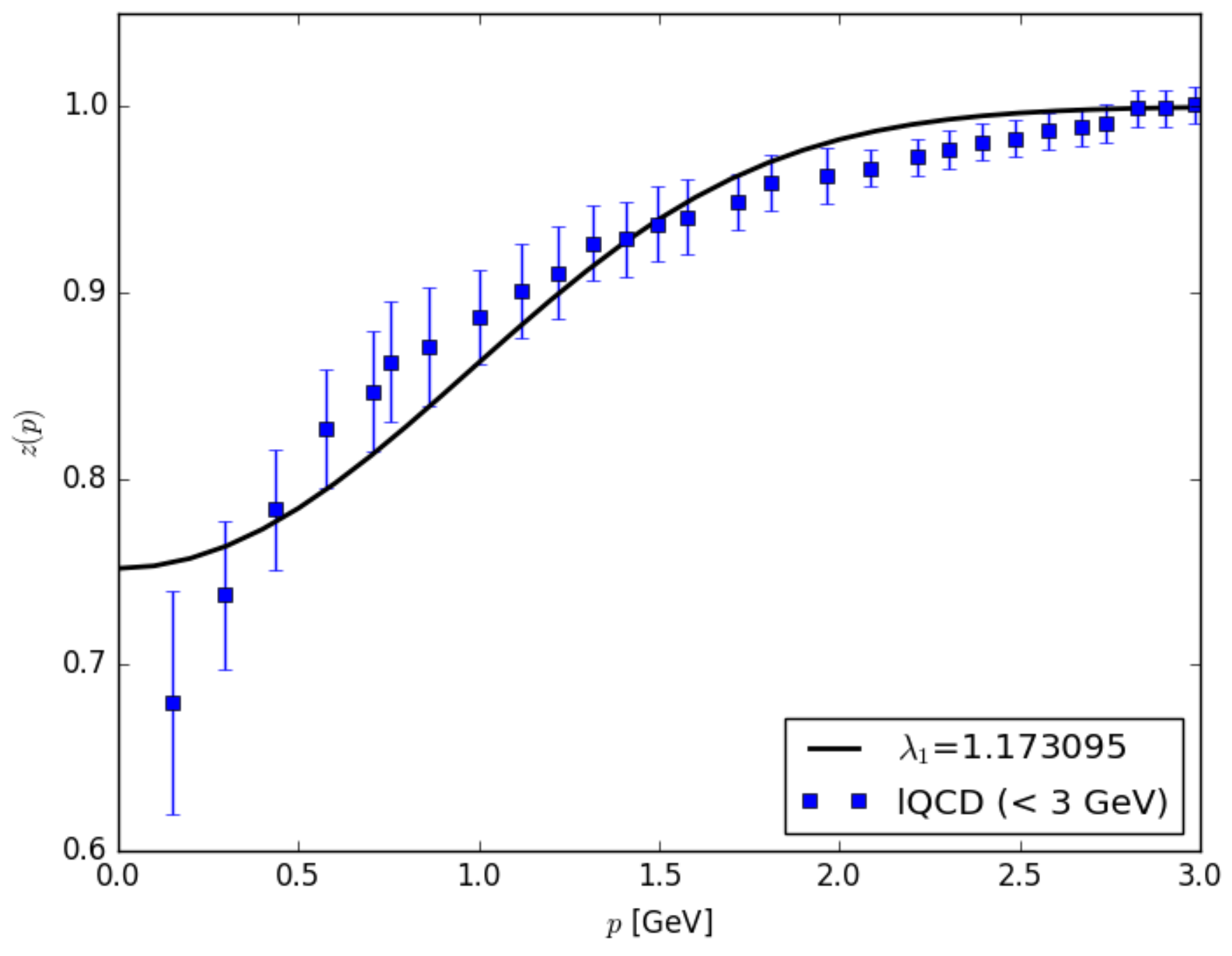}}
\caption{\small{Fit to lattice data from Ref.~\cite{Parappilly:2005ei} for the functions $m(p)$ and $z(p)$, Eq.~(\ref{mz}). 
The fit has been carried out considering results up to 3~GeV.}}
\label{fig:ff}
\end{figure}

Given the form factor functions, it is possible to set the model parameters to reproduce the observed meson phenomenology. 
First, we perform a fit to lQCD results quoted in Ref.~\cite{Parappilly:2005ei} for the functions $m(p)$ and $z(p)$, from which we obtain the values of $\Lambda_0$ and $\Lambda_1$
\begin{eqnarray*}
&&\Lambda_0 = 1092 \pm 22  {\rm \ MeV}\ , \quad
\Lambda_1 = 1173 \pm 60 {\rm \ MeV}\  .
\label{fit}
\end{eqnarray*}
The curves corresponding to the functions $m(p)$ and $z(p)$, together with lQCD data are shown in Fig.~\ref{fig:ff}.

By requiring that the model reproduce the empirical values of three physical quantities, chosen to be the masses of the mesons $\pi$ and $\rho$ and the pion weak decay constant $f_{\pi}$, together with the value of $z(p=0)$, one can determine the model parameters quoted in Table~\ref{tab:param}.
\begin{table}[h]
\begin{center}
\begin{tabular*}{0.2\textwidth}{@{\extracolsep{\fill}} c c }
\hline
\hline
Parameter &  Value \\
\hline
$m$ [MeV] & 2.256
  \\
$G_S$ [GeV$^2$] & 23.30
 \\
$G_V$ [GeV$^2$] & 20.05
  \\
$\kappa_p$ [GeV] & 4.265
  \\
\hline
\hline
\end{tabular*}
\caption{\small{Model parameter values.}}
\label{tab:param}
\end{center}
\end{table}

Regarding the coupling constant $G_0$, we will follow the prescription used in Ref~\cite{Contrera:2012wj}, parameterizing it as $G_0 = \eta\ G_V$.
Therefore, the strength of the isoscalar vector coupling can be evaluated by considering different values for $\eta$, and can be used to tune the model.
Given that the influence of this vector coupling increases with the chemical potential, at zero density $\bar{\omega}$ vanishes for all temperatures, and thus, the vector interactions do not contribute to the mean field thermodynamic potential.

Effective theories that do not include an explicit mechanism of confinement, such as PNJL models, usually present a threshold above which the constituent quarks can be simultaneously on shell.
This threshold, which depends on the model parametrization, is typically of the order of $1$~GeV.
Since the main goal of this work is to describe the properties of vector mesons, particularly the $\rm a_1(1260)$ meson, the parameter set quoted in Table~\ref{tab:param} was chosen in such a way that the threshold, approximately of $1.25$~GeV, is larger than the meson masses obtained within our model. 

Once we have fixed the model parametrization, we can calculate the values of several meson properties. 
Our numerical results are summarized in Table~\ref{tab:prop}, together with the corresponding phenomenological estimates.

In general, it is seen that meson masses, mixing angles and decay constants predicted by the model are in a reasonable agreement with phenomenological expectations.
As in precedent analyses~\cite{Noguera:2008,Hell:2011ic,Carlomagno:2013ona, Carlomagno:2018tyk}, we obtain relatively low values for $m$, and a somewhat large value for the light quark condensate. 
On the other hand, we find that the product $-\langle \bar qq\rangle m$, which gives $7.4\times 10^{-5}$~GeV$^4$ is in agreement with the scale-independent result obtained from the Gell-Mann-Oakes-Renner relation at the leading order in the chiral expansion, namely $-\langle \bar qq\rangle m = f_\pi^2 m_\pi^2/2 \simeq 8.3\times 10^{-5}$~GeV$^4$.

\begin{table}[h]
\begin{center}
\begin{tabular*}{0.35\textwidth}{@{\extracolsep{\fill}} c c c }
\hline
\hline
Parameter &  Value & Empirical \\
\hline
$\bar\sigma_1$ [MeV] & 648.3
 & -  \\
$\bar\sigma_2$ & -0.3307
 & -\\
$\lambda$ [MeV$^{-1}$] & 6.168
 $\times 10^{-4}$  & -\\
$g_{\pi q \overline{q}}$ & 7.068
 & -\\
$g_{\rho q \overline{q}}$ & 4.155
 & -\\
$g_{\rm a_1 q \overline{q}}$ & 3.741
 & -\\
$-\langle \overline{q}q \rangle ^{1/3}$ [MeV] &  329.6
 & 270-330  \\
$m_{\sigma}$ [MeV] & 793.9
 & 400-550  \\
$m_{\rm a_1}$ [MeV] & 1204.1
 & 1260  \\
$f_v$ & 0.1732
 & 0.200 \\
$f_{\rm a}$ & 0.4407
 & 0.230 \\
$\Gamma_{\rho\to \pi\pi}$ [MeV] & 120.7
 & 149.1 \\
$\Gamma_{{\rm a}_1\to \rho\pi}$ [MeV] & 184.7
 & 150-360 \\
\hline
\hline
\end{tabular*}
\caption{\small{Numerical results for various phenomenological quantities.}}
\label{tab:prop}
\end{center}
\end{table}

\subsection{Results at finite temperature} 

We begin this section analyzing the chiral restoration and deconfinement transition at zero chemical potential through the thermal dependence of the corresponding order parameters, namely the chiral quark condensate $\langle \overline{q}q \rangle$ and the trace of the Polyakov loop $\Phi$, respectively.
Both chiral and deconfinement critical temperatures, $T_{ch}$ and $T_\Phi$, are obtained by the position of the peaks in the associated susceptibilities $\chi_{ch}$ and $\chi_\Phi$, defined as
\begin{equation}
\chi_{ch} = \frac{\partial \langle \overline{q}q \rangle }{\partial T}
\quad{\rm and}\quad
\chi_\Phi = \frac{d \Phi} {d T} \ . \nonumber 
\end{equation}

\begin{figure}[h]
\centering
\includegraphics[width=0.45\textwidth]{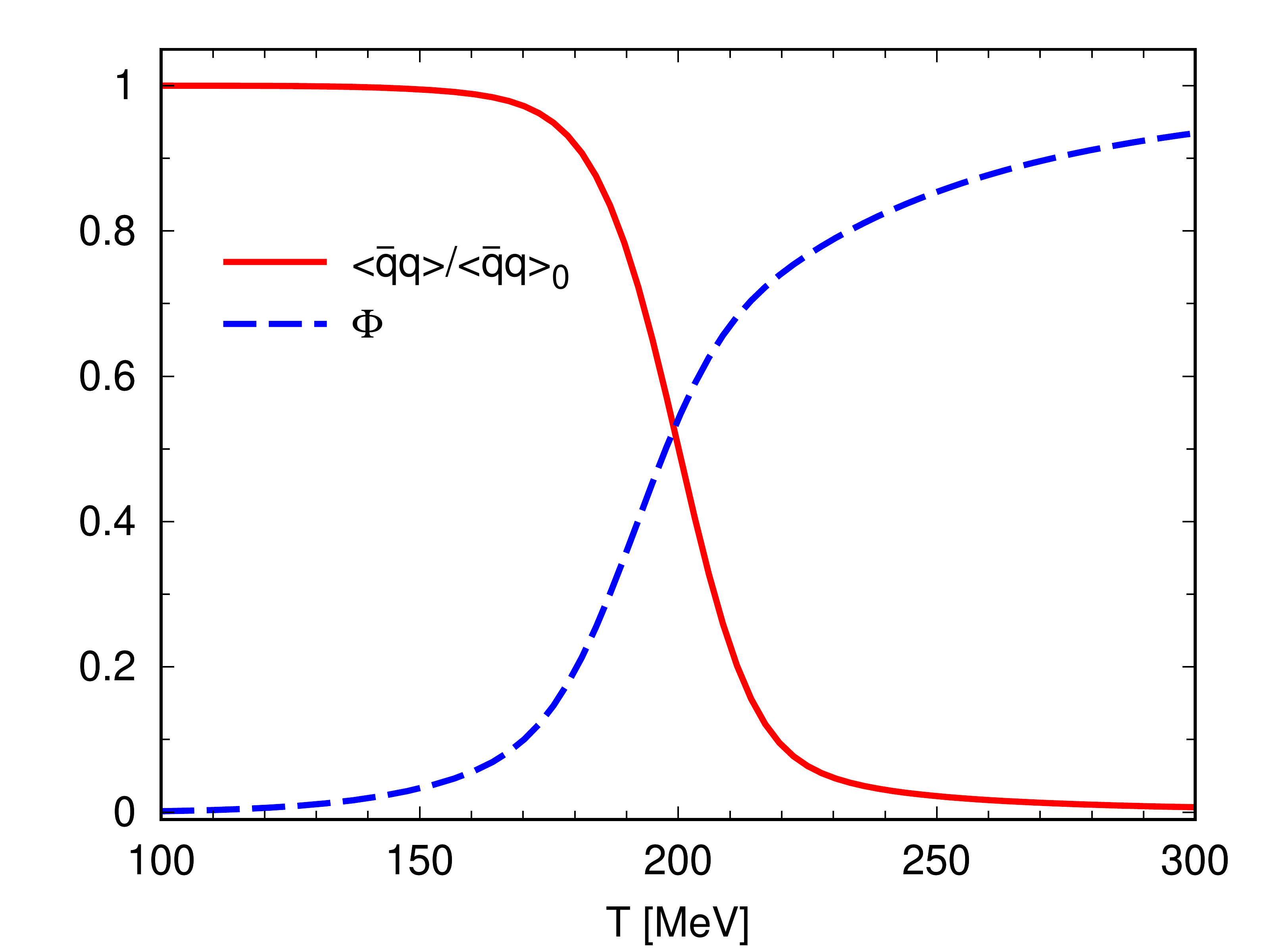}
\caption{\small{Normalized quark condensate and trace of the Polyakov loop in solid and dashed line, respectively, as function of the temperature.}}
\label{fig:o_param}
\end{figure}

In Fig.~\ref{fig:o_param} we plot in solid line the quark condensate normalized by its value at $T=0$, and in dashed line the trace of the Polyakov loop, both as function of the temperature $T$.
Using the potential $\mathcal{U}_{\rm poly}(\Phi,T)$ in Eq.~(\ref{pol}) with $T_0 = 210$~MeV, we find that the chiral and deconfinement critical temperatures are approximately the same (less than $3\%$ of difference).
This behavior is in agreement with lQCD calculations~\cite{Bazavov:2016uvm} and other nlPNJL models~\cite{Contrera:2010kz,Carlomagno:2013ona,Carlomagno:2018tyk}, showing that at $\mu = 0$ chiral restoration and deconfinement take place simultaneously as crossover phase transitions.

The obtained chiral critical temperature, $T_{ch} = 202$~MeV, is somewhat larger than lQCD estimations, $T_{ch}^{lQCD} \sim 160$~MeV~\cite{Bazavov:2016uvm}, however this value is strongly dependent on the presence of the Polyakov loop. 
For instance, nonlocal NJL models (without Polyakov loop)~\cite{Contrera:2007wu,GomezDumm:2005hy,Carlomagno:2015nsa} predict a chiral critical temperature around $120$~MeV, which is quite lower than $T_{ch}^{lQCD}$.
Therefore, in our model framework we can test the effect of the effective Polyakov potential on the chiral critical temperature by modifying the value of $T_0$, staying within the range quoted in Ref.~\cite{Schaefer:2007pw}.

\begin{figure}[h]
\centering
\includegraphics[width=0.45\textwidth]{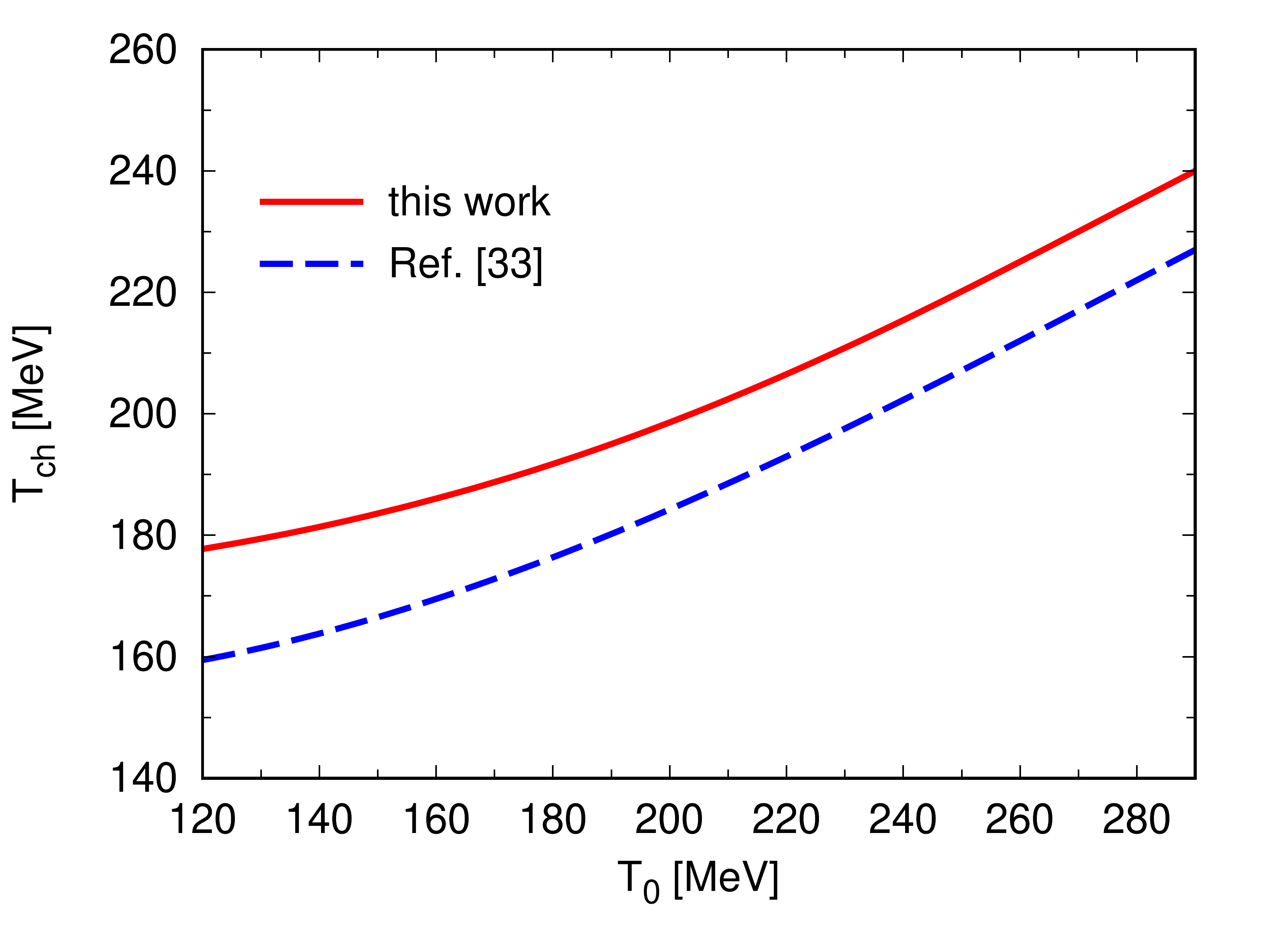}
\caption{\small{Chiral critical temperatures as function of $T_0$ for two different set of parameters. 
In solid line the parametrization used in this work, and in dashed line the one used in Ref.~\cite{IzzoVillafane:2016jnx}.}}
\label{fig:tcrit}
\end{figure}

In Fig.~\ref{fig:tcrit} we plot the chiral critical temperature as a function of $T_0$ for two different set of parameters.
The solid line stands for the parametrization present in Table~\ref{tab:param}, while the dashed one represents the parametrization used in Ref.~\cite{IzzoVillafane:2016jnx}.
With this analysis we can not only estimate how much $T_{ch}$ changes with $T_0$, but also determine its dependence with the parametrization.
We see in the figure that the splitting between lines remains almost constant with a gap of approximately $15$~MeV, and the variation of $T_{ch}$ in the range between $T_0=120-290$~MeV is almost of $70$~MeV.
Hence, with the proper choice of model parameters and $T_0$, we can always obtain a chiral critical temperature in good agreement with lQCD estimations.
Consequently, for all the following figures we will plot our results against the reduced temperature $T/T_{ch}$.

\begin{figure}[h]
\centering
\includegraphics[width=0.45\textwidth]{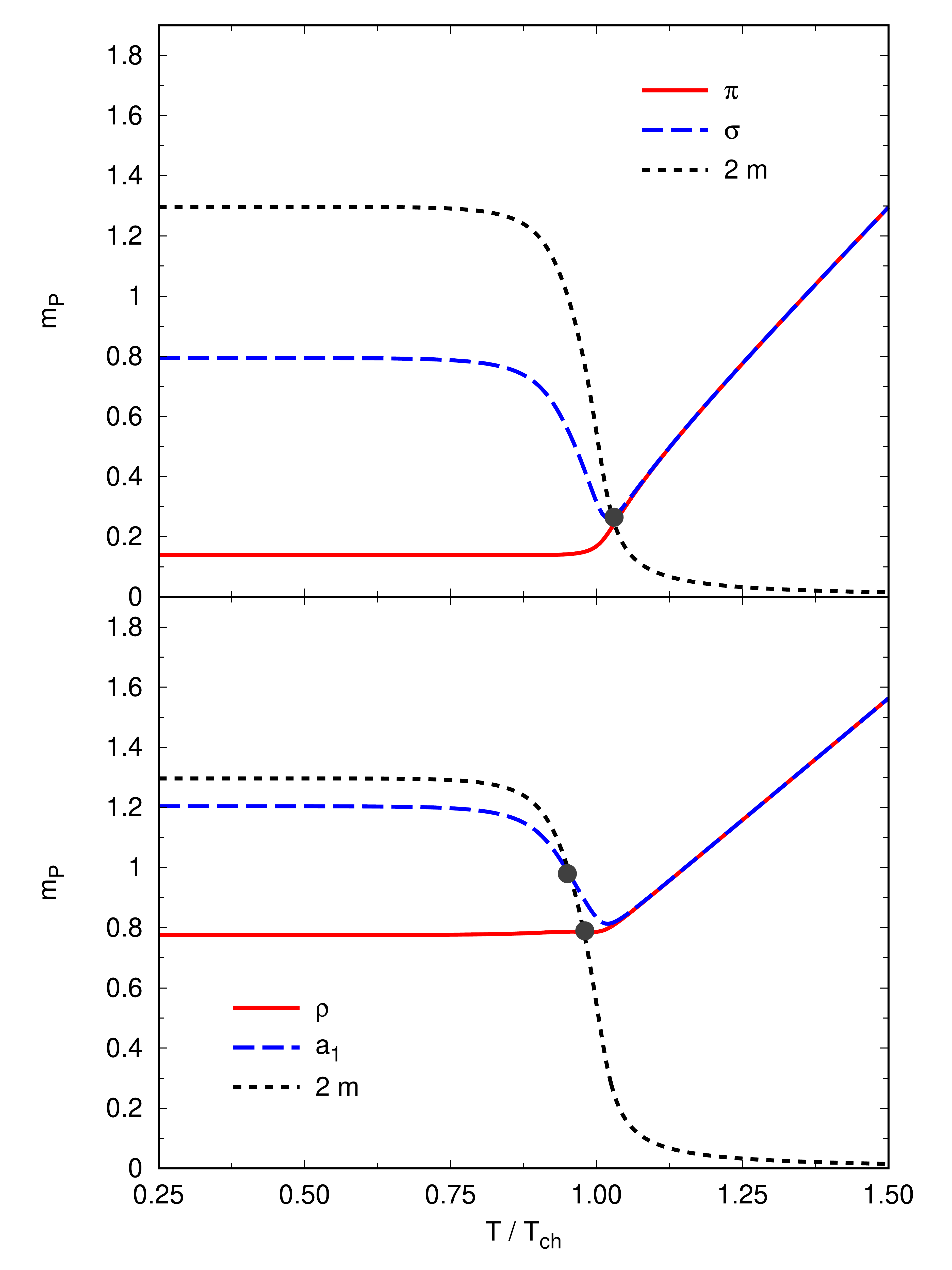}
\caption{\small{Pseudoscalar and scalar (vector and axial-vector) meson masses in solid and dashed line, respectively, in the upper (lower) panel. We also quote, in dotted line, the effective constituent quark mass.}}
\label{fig:masses}
\end{figure}

Our numerical results for the evolution of the meson masses as functions of the reduced temperature are shown in Fig.~\ref{fig:masses}.
The upper panel shows the behavior of the scalar and pseudoscalar mesons $\sigma$ and $\pi$, which are chiral partners.
Similarly, in the lower panel we find the vector and axial vector mesons $\rho$ and a$_1$.
It is seen that pseudoscalar and vector meson masses (solid lines) remain approximately constant up to the critical temperature $T_{ch}$, while scalar and axial-vector meson masses (dashed lines) start to drop somewhat below $T_{ch}$.
Right above $T_{ch}$ masses get increased in such a way that the chiral partners become degenerate, as expected from chiral restoration. 
Afterwards, when the temperature is further increased, the masses rise continuously, showing that they are basically dominated by the thermal energy. 
We also plot in both cases the effective constituent quark masses, shown in dotted lines. 
It is interesting to notice that up to certain temperature $T_m$ (denoted in the figure with a large dot) the mesons have a lower mass than the masses of their constituent quarks. 
However, when $T>T_m$, meson masses are no longer discrete solutions of Eqs.~(\ref{grho}) and~(\ref{gpi}), which implies a passage from the discrete to the continuum, known as Mott transition~\cite{Blaschke:1984yj,Hufner:1996pq}. 
From the figure one can see that the Mott temperature for the scalar and vector mesons is located at $T / T_{ch} \sim 1$. 

\begin{figure}[h]
\centering
\includegraphics[width=0.45\textwidth]{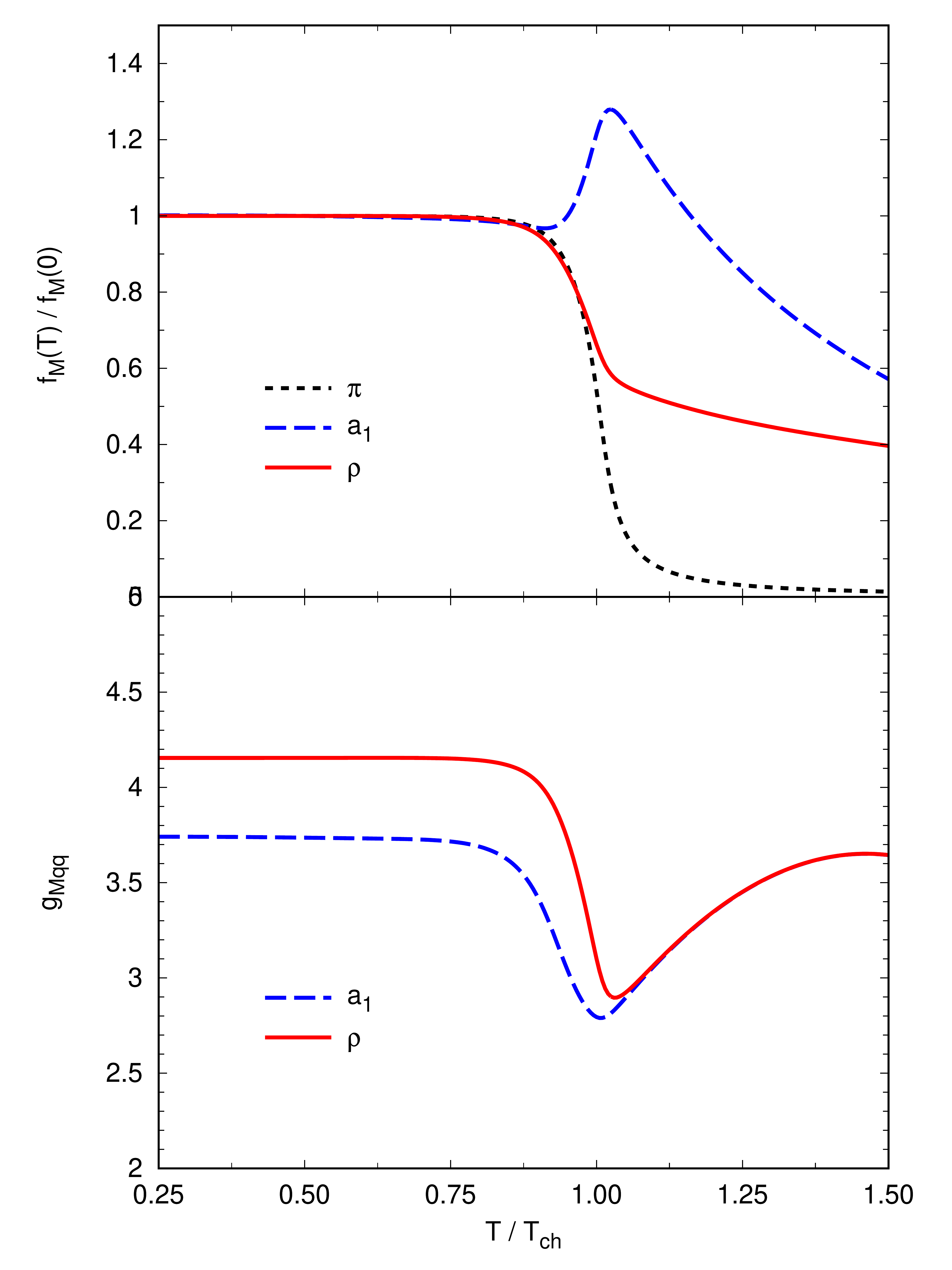}
\caption{\small{Normalized decay constants by its zero temperature value (effective quark meson couplings) in the upper (lower) panel, for the $\rho$ and $\rm a_1$ meson in solid and dashed line, respectively.}}
\label{fig:decay}
\end{figure}

Regarding the thermal behavior of the decay constants, defined in Eqs.~(\ref{efes}) and (\ref{eq:fa}) and plotted in the upper panel of Fig.~\ref{fig:decay}, we see that they remain constant up to near the chiral critical temperature, and beyond this point they tend to zero.
We can observe that the pion decay constant vanishes faster (at $T/T_{ch} \sim 1.5$) than the vector and axial-vector decay constants (at $T/T_{ch} \sim 3$).
Also, $f_{\rm a_1}$ presents a peak immediately above $T_{ch}$ due to the $\rm a_1$ mass behavior, which, before being dominated by the thermal energy, decreases to become degenerated with its chiral parter.

The lower panel of Fig.~\ref{fig:decay} shows the quark-meson couplings of the vector and axial-vector mesons in solid and dashed line, respectively.
Both quark-meson couplings present a similar thermal behavior, remaining constant at low temperatures and decreasing before reaching $T_{ch}$.
Near above the critical temperature, the couplings become degenerated. 

\begin{figure}[h]
\centering
\includegraphics[width=0.45\textwidth]{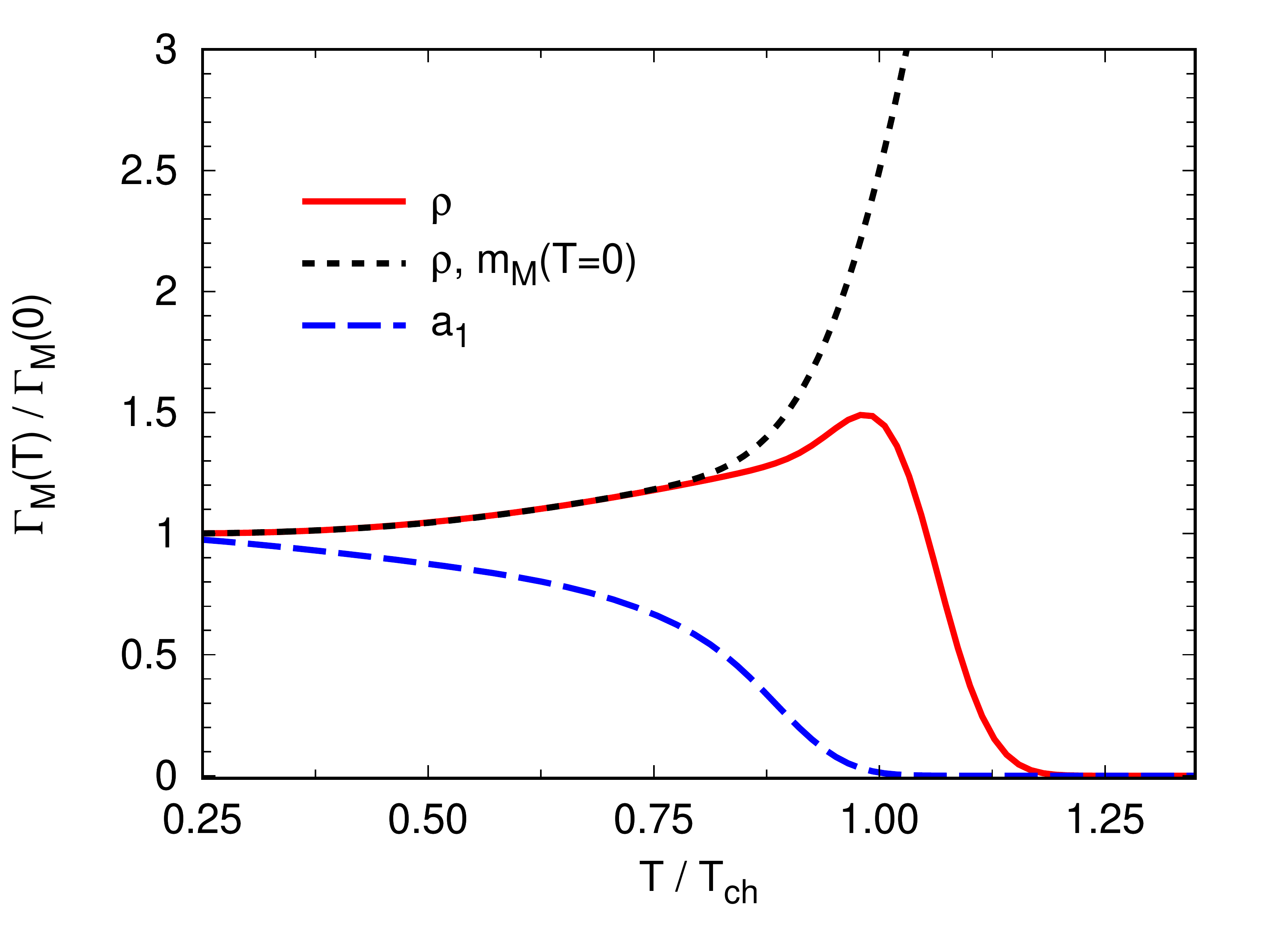} 
\caption{\small{In solid and dashed lines the $\rho$ and $\rm a_1$ decay widths, respectively, as function of the reduced temperature. In dotted line, $\Gamma_\rho$ for constant meson masses.}}
\label{fig:gammas}
\end{figure}

In Fig.~\ref{fig:gammas} we plot the $\rho$ and $\rm a_1$ decay widths, $\Gamma_\rho$ and $\Gamma_{\rm a_1}$, defined in Eq.~(\ref{gammaofT}). 
It can be seen that the solid line, which represents $\Gamma_\rho$, and the dashed line, $\Gamma_{\rm a_1}$, show a decreasing behavior as $T$ rises.
It is found in these cases that the thermal dependence in the meson masses directly affects the decay widths.
This happens because of the kinematic condition in Eq.~(\ref{gammaofT}), which goes to zero as $T$ increases. 
As a result the decay widths tend to vanish, even when the phase space becomes larger due to the Bose enhancement.
On the contrary, if the masses are considered to be constant for all temperatures, then $\Gamma_\rho$ would follow the dotted line curve in Fig.~\ref{fig:gammas}, and monotonously increase as a function of $T$.
This is fully consistent with the results obtained, for instance, within QCD sum rules~\cite{Ayala:2012ch}, where the same condition is imposed to the masses.

Despite having the same general tendency towards zero, $\Gamma_\rho$ and $\Gamma_{\rm a_1}$ present a transition at different temperatures.
In the case of the $\rho$ meson, is only when $T>T_{ch}$ that $\Gamma_\rho$ starts to drop, since the $m_\pi$ grows faster than the $m_\rho$ with temperature (see Fig.~\ref{fig:masses}).

For the $\rm a_1$ decay however, the width begins to diminish before the chiral critical temperature, vanishing close to $T_{ch}$.
This is caused by the chiral partners mass degeneration, near above $T_{ch}$ the vector and axial-vector mesons have approximately the same mass (see lower panel of Fig.~\ref{fig:masses}), and therefore when this happens the kinematic condition in Eq.~(\ref{gammaofT}), vanishes.

Concerning the axial-vector decay width $\Gamma_{\rm a_1}$ it begs to be mentioned that, while we are only considering the main channel of the partial decay $\rm a_1 \rightarrow \rho \pi$, the physical decay process is actually $\rm a_1 \rightarrow \pi \pi \pi$.
Other partial widths contribute approximately with $40 \%$ of the total width~\cite{Tanabashi:2018oca}.
Moreover, the decay $\rm a_1 \rightarrow \sigma \pi$ and the direct decay $\rm a_1 \rightarrow \pi \pi \pi$ are also kinematically allowed for a higher temperature range, since $m_\sigma$ and $m_\pi$ are smaller than $m_\rho$ near and above $T_{ch}$. 
Even though these processes are not calculated here, they would contribute to the total width and could modify the decreasing behavior of $\Gamma_{\rm a_1}$.

\section{QCD phase diagram} 
\label{QCDpd}

Through the study of the order parameters one can find regions in which the chiral symmetry is either broken or approximately restored through first order or crossover phase transitions, and phases in which the system remains either in confined or deconfined states. 

For relatively high temperatures chiral restoration takes place as a crossover, whereas at low temperatures the order parameter has a discontinuity at a given critical chemical potential signaling a first order phase transition. 
This gap in the quark condensate induces also a jump in the trace of the PL.
The value of $\Phi$ at both sides of the discontinuity indicate if the system remains confined or not.
Values close to zero or one correspond to confinement or deconfinement, respectively.

When the chiral restoration occurs as a first order phase transition, the PL susceptibility present a divergent behavior at the chiral critical temperature even when the order parameter $\Phi$ remains close to zero.
Therefore, another definition is needed for the deconfinement critical temperatures in this region of the phase diagram. 
As in Ref~\cite{Contrera:2010kz}, we define the critical temperature requiring that $\Phi$ takes a value in the range between $0.4$ and $0.6$, which could be taken as large enough to denote deconfinement.

Given that deconfinement occurs at temperatures where the order parameter becomes close to one, the phase in which quarks remain confined (signaled by $\Phi\lesssim 0.4$) even though chiral symmetry has been restored is usually referred to as a quarkyonic phase~\cite{McLerran:2007qj,McLerran:2008ua,Abuki:2008nm}.

As it is explained in Ref~\cite{Contrera:2012wj} the isoscalar vector coupling constant $G_0$ is considered to be a free parameter which may be adjusted in the MFA to reproduce the behavior of thermodynamic properties obtained in lattice QCD and other effective theories.
Therefore, as was stated, the vector coupling strength will be evaluated by defining the ratio $\eta = G_0/G_V$, and to study how the vector interactions affect the transitions and the location of the CEP, we built the corresponding phase diagrams for different values of $\eta$.

\begin{figure}[h]
\centering
\includegraphics[width=0.49\textwidth]{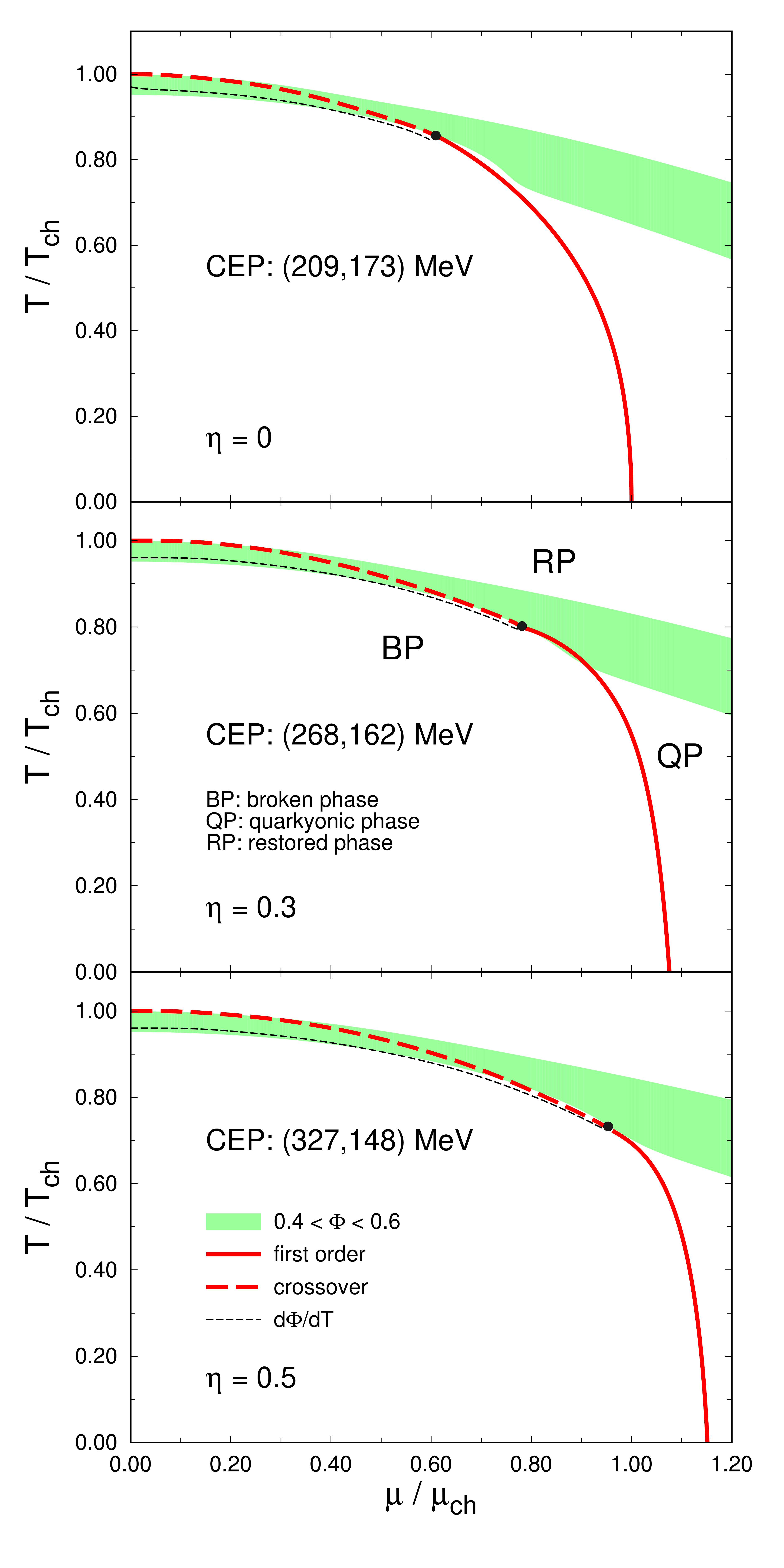}
\caption{\small{QCD phase diagrams for the polynomial PL potential within a nlPNJL model for $\eta=0,0.3,0.5$ in the upper, middle and lower panel, respectively.}}
\label{fig:QCDpd}
\end{figure}

In Fig.~\ref{fig:QCDpd} the first order phase transition for the chiral symmetry restoration can be seen in solid lines, while the dashed lines show the crossover transition. 
In addition, the deconfinement transition range, defined by $0.4<\Phi<0.6$, is denoted with the color shaded area.
Finally, the dot indicates the position of the critical endpoint. 

If we move in the $T-\mu$ plane, along the first order phase transition curve, the critical temperature rises from zero up to a critical endpoint (CEP) temperature $T_{CEP}$, while the critical chemical potential decreases from $\mu_{ch}$ to a critical endpoint chemical potential $\mu_{CEP}$. 
Beyond this point, the chiral restoration phase transition proceeds as a crossover.

In Table~\ref{tab:cep} we summarize, for the three considered values of $\eta$, the critical temperatures $T_c$, critical chemical potentials $\mu_c$ and the CEP coordinates ($\mu_{\rm CEP}$, $T_{\rm CEP}$).
\begin{table}[h]
\begin{center}
\begin{tabular*}{0.4\textwidth}{@{\extracolsep{\fill}} cccc }
\hline 
\hline 
$G_0$ & 0 & 0.3\ $G_V$ & 0.5\ $G_V$\\
\hline
$T_{\rm CEP}$ [MeV] & 173 & 162 & 148  \\
$\mu_{\rm CEP}$ [MeV] & 209 & 268 & 327 \\
\hline
$T_c$ [MeV]  & 202 & 202 & 202 \\
$\mu_c$ [MeV]  & 343 & 369 & 395\\
\hline 
\hline 
\end{tabular*}
\caption{\small{CEP coordinates and critical temperatures and densities for the different cases of vector strength.}}
\label{tab:cep}
\end{center}
\end{table}

In the upper panel of Fig.~\ref{fig:QCDpd} we plot the phase diagram for a mean field theory without vector interactions ($\eta =0 $), while in the two lower panels we show the phase diagrams in presence of the isoscalar vector coupling for two different values of $\eta$, $\eta=0.3$ and $\eta=0.5$. 
We notice that the influence of the vector interaction increases with
the chemical potential, in particular, the position of the CEP and the values of $\mu_{ch}$ reflect notably this influence (for increasing $\eta$ the CEP's tend to be located towards a lower $T$ and higher $\mu$).

In addition, from Fig.~\ref{fig:QCDpd} two different behaviors for the hadronic matter depending on the chemical potential are observed.
When the chiral restoration transition is first order ($\mu > \mu_{CEP}$), 
as the temperature increases we find a transition from a hadronic phase with broken chiral symmetry (BP), to a quarkyonic phase (QP) where the chiral symmetry is restored but the quarks are still confined into hadrons. 
If the temperature continues raising, the deconfinement transition takes place reaching a partonic phase in which quarks are deconfined and the chiral symmetry is restored (RP).
On the other hand, for densities lower than $\mu_{CEP}$ one goes from the BP to the RP though crossover phase transitions when the temperature increases. Hence, the chiral restoration and deconfinement take place almost simultaneously.

\section{Summary and conclusions}
\label{summary}

In this work we have analyzed the thermal properties of the lightest vector mesons and characterized the chiral and deconfinement transitions at finite $T$ and $\mu$, within the context of a $SU(2)$ nonlocal NJL model with vector interactions and wave function renormalization (WFR).
Gauge interactions have been effectively introduced through a coupling between quarks and a constant background color gauge field, the Polyakov field, whereas gluons self-interactions have been implemented through the polynomial effective Polyakov loop potential.

Regarding the vacuum properties, we have been able to find a parameter set that reproduces lattice QCD results for the momentum dependence of the effective quark mass and WFR, and the related threshold, where constituent quarks can be simultaneously on shell, is larger than the obtained meson masses.

At finite temperature, we study the masses, widths and decay constants of the lightest vector and axial-vector mesons.
We found, as expected, that meson masses and decay constants remain approximately constant up to the critical chiral temperature.
Beyond the chiral critical temperature, the masses get increased, becoming degenerated with their chiral partners. 

For the process $\rho \rightarrow \pi \pi$, the decay width starts to drop above the chiral critical temperature, since beyond this temperature the $\pi$ mass grows faster than the $\rho$ mass.

For the ${\rm a}_1$ decay, due to the chiral partners mass degeneration, the width begins to diminish before the chiral critical temperature, vanishing close to $T_{ch}$. 
This indicates that the non considered decay processes for the ${\rm a}_1$ could be relevant for temperatures close to the chiral critical temperature.

At zero $\mu$, the model shows a crossover phase transition, corresponding to the restoration of the $SU(2)$ chiral symmetry. 
In addition, one finds a deconfinement phase transition, which occurs at almost the same critical temperature.

On the other hand, at zero temperature chiral restoration takes place via a first order transition.

At finite density, in the first order region, the critical temperatures for the restoration of the chiral symmetry and deconfinement transition begin to separate. The region between them denotes the quarkyonic phase.

In addition, when the vector interactions are added to the model ($\eta \neq 0$) we found that the position of the CEP and $\mu_{ch}$ are influenced by the strength of the coupling constant $G_0$, noticing that when $\eta$ increases the critical end-point appears at a lower $T$ and higher $\mu$.

\section*{Acknowledgements}

This work has been partially funded by the National University of La Plata, Project No.\ X824, by CONICET under Grant No. PIP 2017/1220170100700 and by  ANPCyT, Grant No. PICT 2017-0571. 

The authors would like to thank D.~G\'omez~Dumm and N.N.~Scoccola for useful discussions.

\begin{widetext}

\appendix

\section{Screening masses and the decays constants}
\label{App_masses_decays}

Here we quote the analytical expressions for the effective thermal meson propagators in the imaginary time formalism, $G_{M}(\nu_m,\vec{p})$, where $\nu_m = 2 m \pi T$ are the bosonic Matsubara frecuencies. 
Also we will present the functions defined in Eq.~(\ref{efes}) to calculate the decay constants $f_\pi$ and $f_v$ (see Ref.~\cite{Villafane:2016ukb} for details of their derivations).

For the vector and axial vector sector, the functions $G_\rho(\nu_m,\vec{p})$ and $G_{{\rm a}_1}(\nu_m,\vec{p})$ can be written as
\begin{eqnarray}
 G_{\rho \choose {\rm a}_1}(\nu_m,\vec{p}) &=& \dfrac{1}{G_V}- 8\!\! 
\sum_{c=r,g,b} T \!\! \sum_{n=-\infty}^{\infty} \int \frac{d^3\vec q}{(2\pi)^3}  
                               \, h^{2}(q_{nc})\,
                               \dfrac{z(q_{nc}^+)z(q_{nc}^-)}{D(q_{nc}^{+})D(q_{nc}^{-})} 
                         \left[\dfrac{q_{nc}^{2}}{3}+\dfrac{2(p_m\cdot q_{nc})^{2}}{3p^{2}}-
                               \dfrac{p_m^{2}}{4}\pm m(q_{nc}^{-})m(q_{nc}^{+})\right]\ , \nonumber\\
\label{grho} 
\end{eqnarray}
with $D(q_{nc}) = q_{nc}^2 + m^2(q_{nc})$ and $q_{nc}^\pm = q_{nc} \pm p_m/2$.

In order to calculate the physical state $\delta \tilde{\vec \pi}$ we define from Eq.~(\ref{eq:quad}) the mixing function $\lambda(p^2)$ as
\begin{equation}
\lambda(p^2) = \dfrac{G_{\pi a}(p^2)}{L_-(p^2)} \ .
\end{equation}
Thus, to calculate the pion mass we find
\begin{equation}
G_{\tilde{\pi}}(p_m^2)= G_{\pi}(p_m^2)- \lambda(p_m^2)\ G_{\pi a}(p_m^2)\,p_m^2\ ,
\label{gpi}
\end{equation}
with
\begin{eqnarray}
G_{\pi}(p_m^2) & = & \dfrac{1}{G_S} \, - \, 8 
		\sum_{c=r,g,b} T \sum_{n=-\infty}^{\infty} \int \frac{d^3\vec q}{(2\pi)^3} \
		 \ g(q_{nc})^2\,
                	\dfrac{z(q_{nc}^+)z(q_{nc}^-)}{D(q_{nc}^{+})D(q_{nc}^{-})} 
              \Big[(q_{nc}^{+}\cdot q_{nc}^-)\,+\,m(q_{nc}^{+})\,m(q_{nc}^{-})\Big] \ , \nonumber \\
G_{\pi a}(p_m^2) & = & \dfrac{8}{p_m^{2}}\, 
		\sum_{c=r,g,b} T \sum_{n=-\infty}^{\infty} \int \frac{d^3\vec q}{(2\pi)^3} \ 
		\ g(q_{nc})^2\,
                 	\dfrac{z(q_{nc}^+)z(q_{nc}^-)}{D(q_{nc}^{+})D(q_{nc}^{-})}
                  \Big[(q_{nc}^{+}\cdot p_m)\,m(q_{nc}^{-})-(q_{nc}^{-}\cdot
                 p_m)\,m(q_{nc}^{+})\Big] \ , \nonumber \\
L_{-}(p_m^{2}) & = &\dfrac{1}{G_V}\, - \, 8
		\sum_{c=r,g,b} T \sum_{n=-\infty}^{\infty} \int \frac{d^3\vec q}{(2\pi)^3} \ 
		\ g(q_{nc})^2\,
                 	\dfrac{z(q_{nc}^+)z(q_{nc}^-)}{D(q_{nc}^{+})D(q_{nc}^{-})} 
                 \left[q_{nc}^{2}-\dfrac{2(p_m\cdot q_{nc})^{2}}{p_m^{2}}+\dfrac{p_m^{2}}{4} -
                	 m(q_{nc}^{-})m(q_{nc}^{+})\right] . \nonumber\\
                 \label{lpm}                 
\end{eqnarray}

In the case of the $f_\pi$ and $f_v$ decay constants we need to evaluate the functions $F_0(p_m^2)$, $F_1(p_m^2)$ and $J_{V}^{\rm I,II}(p_m^2)$. 
After a rather lengthy calculation we find for these functions the following expressions
\begin{eqnarray}
F_0 (p_m^2) &=& 8\sum_{c=r,g,b} T \sum_{n=-\infty}^{\infty} \int \frac{d^3\vec q}{(2\pi)^3} \ g(q_{nc})\, \dfrac{z(q_{nc}^+)z(q_{nc}^-)}{D(q_{nc}^+)D(q_{nc}^-)}\ 
             \left[ (q_{nc}^+\cdot q_{nc}^-) + m(q_{nc}^+)\,m(q_{nc}^-)\right] \ ,\nonumber\\
F_1 (p_m^2) &=& 8\sum_{c=r,g,b} T \sum_{n=-\infty}^{\infty} \int \frac{d^3\vec q}{(2\pi)^3} \ g(q_{nc})\, \dfrac{z(q_{nc}^+)z(q_{nc}^-)}{D(q_{nc}^+)D(q_{nc}^-)}\ 
             \left[ (q_{nc}^+\cdot p_m)\,m(q_{nc}^-) - (q_{nc}^-\cdot p_m)\,m(q_{nc}^+)\right]\ ,
\label{f01}
\end{eqnarray}
and
\begin{eqnarray}
J_V^{\rm (I)}  (p_m^2) &=& -\,4\sum_{c=r,g,b} T \sum_{n=-\infty}^{\infty} \int \frac{d^3\vec q}{(2\pi)^3} \ g(q_{nc})\,\Bigg\lbrace
              \dfrac{3}{2}\,\dfrac{[z(q_{nc}^+)+z(q_{nc}^-)]}{D(q_{nc}^+)D(q_{nc}^-)}\Big[
              (q_{nc}^+\cdot q_{nc}^-) + m(q_{nc}^+)\,m(q_{nc}^-) \Big] \nonumber \\
              & & + \; \dfrac{1}{2}\,\dfrac{z(q_{nc}^+)}{D(q_{nc}^+)} \, +
                  \, \dfrac{1}{2}\, \dfrac{z(q_{nc}^-)}{D(q_{nc}^-)}\, + \,\dfrac{q_{nc}^2}{(q_{nc}\cdot p_m)}
                  \left[\dfrac{z(q_{nc}^+)}{D(q_{nc}^+)} - \dfrac{z(q_{nc}^-)}{D(q_{nc}^-)}\right]\nonumber \\
              & & + \,\dfrac{z(q_{nc}^+)z(q_{nc}^-)}{D(q_{nc}^+)D(q_{nc}^-)}\,
              \left[(q_{nc}\cdot p_m) - \dfrac{q_{nc}^2\,p_m^2}{(q_{nc}\cdot p_m)}\right]\,
                  \bigg[-\,\bar\sigma_1\, \big[m(q_{nc}^+) + m(q_{nc}^-)\big]\,\alpha^+_g(q_{nc},p_m) \nonumber \\
              & & + \; \bar\sigma_2\,\big[q_{nc}^2 + \dfrac{p_m^2}{4}
                  - m(q_{nc}^+)\,m(q_{nc}^-) \big]\,\alpha^+_f (q_{nc},p_m)\,\bigg] \Bigg\rbrace\ , \nonumber\\
J_V^{\rm (II)} (p_m^2) &=& -\,4\sum_{c=r,g,b} T \sum_{n=-\infty}^{\infty} \int \frac{d^3\vec q}{(2\pi)^3} \
              \dfrac{z(q_{nc})}{D(q_{nc})}\left\lbrace\dfrac{q_{nc}^2}{(q_{nc}\cdot p_m)}
              \Big[g(q_{nc}^+)-g(q_{nc}^-)\Big] \right. \nonumber\\ 
              & & + \left. \left[(q_{nc}\cdot p_m) -
              \dfrac{q_{nc}^2\,p_m^2}{(q_{nc}\cdot p_m)}\right]\alpha^+_g (q_{nc},p_m)
              \right\rbrace \ ,
\end{eqnarray}
where
\begin{equation}
\alpha^+_f (q,p) \ = \ \int_{-1}^1 d\lambda \,\dfrac{\lambda}{2}\,
                   f^\prime\left( q-\lambda \dfrac{p}{2}\right) \ .
\end{equation}

\section{Analytic expressions for the decays widths}
\label{App_widths}

We start this Appendix with the analytical expression for the factors $\tilde G_{\rho\pi\pi}(p_m^2,q_{1,m}^2,q_{2,m}^2)$, and $\tilde F_i (p_m^2,q_{1,m}^2,q_{2,m}^2)$ 
\begin{eqnarray}
\tilde G_{\rho\pi\pi} (p_m^2,q_{1,m}^2,q_{2,m}^2) & = & Z_\rho^{1/2}\,Z_\pi\,
\bigg[ G_{\rho\pi\pi}(p_m^2,q_{1,m}^2,q_{2,m}^2) + 
 \lambda(q_{2,m}^2) \ G_{\rho\pi a} (p_m^2,q_{1,m}^2,q_{2,m}^2) \nonumber \\
 && + \lambda(q_{2,m}^2)^2 \
G_{\rho a a} (p_m^2,q_{1,m}^2,q_{2,m}^2) \bigg] \ , \nonumber \\
\tilde F_i (p_m^2,q_{1,m}^2,q_{2,m}^2) & = &  Z_\rho^{1/2}\,Z_\pi^{1/2}\,Z_{\rm a_1}^{1/2} \  F_i (p_m^2,q_{1,m}^2,q_{2,m}^2) \ .
\label{grhopipi}
\end{eqnarray}

To calculate the $\rho\to\pi\pi$ decay amplitude, we have to evaluate the functions $G_{\rho x y}(p_m^2,q_{1,m}^2,q_{2,m}^2)$, where subindices $x$ and $y$ stand for either $\pi$ or $a$, at $q_1^2 = q_2^2 = (p-q_1)^2 = -m_\pi^2$, $p^2 = -m_\rho^2$. 
It is convenient to introduce the momentum $v = q_1 - p/2$, which satisfies $p\cdot v = 0$, $v^2 = m_\rho^2/4 -m_\pi^2$. 
Then
\begin{eqnarray}
G_{\rho x y}(p_m^2,q_{1,m}^2,q_{2,m}^2) & = & 16 \sum_{c=r,g,b} T \sum_{n=-\infty}^{\infty} \int \frac{d^3\vec q}{(2 \pi)^3}
\,g(q_{nc})\,g\left(q_{nc} + v_m/2 + p_m/4\right)\,g\left(q_{nc} + v_m/2 -
p_m/4\right) \nonumber \\ 
&& \times \dfrac{z(q_{nc}^+)z(q_{nc}^-)z(q_{nc}+v_m)}{D(q_{nc}^+)D(q_{nc}^-)D(q_{nc}+v_m)}
\; f_{x y}(q_{nc},p_m,v_m) \ ,
\label{grhopipiint}
\end{eqnarray}
where we have defined $q_{nc}^\pm = q_{nc} \pm p_m/2$. 
We find for $f_{xy}(q,p,v)$ the expressions (to simplify the notation we will omit the subindexes in $q$, $p$ and $v$)
\begin{eqnarray}
f_{\pi\pi} &=& \bigg[(q^+ \cdot q^-) + m(q^+)\,m(q^-)\bigg] \, \bigg[1 + \dfrac{(q\cdot v)}{v^2}\bigg]
\nonumber \\
& & -\, \dfrac{(q\cdot v)}{v^2} \bigg\{ 2\, \Big[\,q\cdot
(q+v)\Big] \, + \, m(q+v)\,\Big[m(q^+)+\,m(q^-)\Big]\bigg\} \ ,\nonumber \\
f_{\pi a} &=& - 2\, m(q+v) \left[ (q^+\cdot q^-)\, - \,2\, \dfrac{(q\cdot v)^2}{v^2}
\, + \, m(q^+)m(q^-)\right] \nonumber \\
& & + \, \bigg[1 + \dfrac{(q\cdot v)}{v^2}\bigg] \, \bigg\{(q^+\cdot p)\,m(q^-)
- (q^-\cdot p)\,m(q^+)\,-\,2(q \cdot v) \Big[m(q^+)+m(q^-)\Big]\bigg\} \ ,
\nonumber \\
f_{aa} &=&
\bigg[1 + \dfrac{(q\cdot v)}{v^2} \bigg] \bigg[q^{+2}\,q^{-2}\, - \, (q^+\cdot
q^-) \, (q+v)^2 \, - \, \Big( v^2 + \dfrac{p^2}{4} \Big) m(q^+)m(q^-) \bigg]
\nonumber \\
& & + \; m(q+v) \bigg\{ m(q^+)\, (q^- \cdot p)\, -\, m(q^-)\, (q^+ \cdot p)\,
+\, \dfrac{(q\cdot v)}{v^2}\bigg(v^2 - \dfrac{p^2}{4} \bigg)\,
\Big[ m(q^+)\, + \, m(q^-)\Big] \bigg\} \nonumber \\
& & + \; 2\,\dfrac{(q\cdot v)}{v^2} \,(q+v)^2 \bigg[(q\cdot v) - \dfrac{p^2}{4}
\bigg]\ .
\end{eqnarray}

In the case of the ${\rm a}_1\to\rho\pi$ decay amplitude the functions $\tilde F_i (p^2,q_1^2,q_2^2)$ can be written as
\begin{eqnarray}
 F_i (p_m^2,q_{1,m}^2,q_{2,m}^2)  &=&
     16\, \sum_{c=r,g,b} T \sum_{n=-\infty}^{\infty} \int \frac{d^3\vec q}{(2 \pi)^3}
		\,g(q_{nc})\,g (q_{nc} + q_{1,m}/2 )\,g (q_{nc} - q_{2,m}/2) \nonumber \\
		&& \times 	\dfrac{z(q_{nc}^+)z(q_{nc}^-)z(\bar q_{nc})}{D(q_{nc}^+)D(q_{nc}^-)D(\bar q_{nc})}\ 
					f^i(p_m^2,q_{1,m}^2,q_{2,m}^2) \ ,  
\end{eqnarray}
where we have defined $q_{nc}^\pm = q_{nc} \pm (q_{1,m}+q_{2,m})/2$ and $\bar{q_{nc}} = q_{nc} + (q_{1,m}-q_{2,m})/2$. 
We obtain
\begin{eqnarray}
    f^1 &=&  m(\bar q)\, q^+ \cdot q^- \ - \  m(q^+)\, \bar{q}\cdot q^- \ - \ m(q^-)\, \bar{q}\cdot q^+ \ - \
		       m(\bar q)m(q^+)m(q^-) \ + \ 2\, \beta_a\, \Big(m(q^+)-m(\bar q)\Big) \nonumber\\
		  & &  + \ \lambda(q_2^2) \, \Bigg[m(\bar q)m(q^-)\, q^+ \cdot p_2 \ + \
						    m(\bar q)m(q^+)\, q^- \cdot p_2 \ - \
		                                    m(q^+)m(q^-)\, \bar q \cdot p_2 \ - \
		                                    q^{+2}\, (q^- \cdot p_2) \nonumber\\
		                                    & & + \
		                                    ( 2\, q \cdot p_2 \ + \ p_1 \cdot p_2 )  (q^+ \cdot q^-) \ + \
		                                    2\,\beta_a\,\Big(\bar q^2 \ - \ q^{+2}\Big) \Bigg] \\
    f^3 &=&  \frac{m(\bar q)}{2} \ +\ \frac{m(q^+)}{2} \ - \ 2\, \alpha_b\, m(q^+) \ + \
				2\, \beta_c\, \Big(m(q^+) \ - \ m(\bar q)\Big) \nonumber\\
				& & + \ \lambda(q_2^2)\, \left[m(\bar q)m(q^+) \ - \ \dfrac{\bar q^2\ + \ q^{+2}}{2} \ - \ 
				 2\, \alpha_b \, \Big(m(\bar q)m(q^+) \ - \ q^{+2} \Big)\ 
				 + \ 2\, \beta_c\, \Big(\bar q^2 \ - \ q^{+2}\Big)\right] \\
    f^4 &=& \frac{m(\bar q)}{2} \ +\ \frac{m(q^-)}{2} \ + \ \alpha_a\, \Big(m(q^-) \ - \ m(q^+)\Big) \ + \
				2\, \beta_d\, \Big(m(q^+) \ -\ m(\bar q)\Big) \nonumber\\
				& & + \ \lambda(q_2^2)\, \Bigg\{
			    \frac{1}{2}\Big[m(q^-)\big(m(\bar q) \ + \ m(q^+)\big) \ +\ m(\bar q)m(q^+) \ - \ \bar q^2 \Big]
			    \ + \ 2\, \beta_d\, \Big(\bar q^2 \ - \ q^{+2}\Big)  \nonumber\\
			    & & \hspace{1.5cm} + \ \alpha_a\, \Big[m(q^-)\big(m(\bar q) \ + \ m(q^+)\big) \ - \ 
			    m(\bar q)m(q^+) \ + \ q^{+2} \Big] \Bigg\} \ ,
\end{eqnarray}
where the coefficients $\alpha_i$ and $\beta_i$ are
\begin{equation}
    \alpha_a \ = \ \dfrac{(p_1\cdot q)(p_1\cdot p_2)\ - \ (p_2\cdot q)p_2^2}{(p_1\cdot p_2)^2 \ - \ p_1^2p_2^2} 
	\qquad\qquad\qquad\qquad
    \alpha_b \ = \ \dfrac{(p_2\cdot q)(p_2\cdot p_1)\ - \ (p_1\cdot q)p_1^2}{(p_1\cdot p_2)^2 \ - \ p_1^2p_2^2}
\end{equation}
\begin{eqnarray}
  \beta_a & = & \dfrac{q^2 (p_1\cdot p_2)^2 \ + \ p_1^2 (p_2\cdot q)^2 \ + \ p_2^2 (p_1\cdot q)^2  \ - \
	  2 \,(p_1\cdot q)(p_2\cdot q)(p_1\cdot p_2) \ - \ q^2 p_1^2 p_2^2 }{2\, \big[(p_1\cdot p_2)^2 \ - \ p_1^2p_2^2\big]} 
	\nonumber \\
   \beta_b & = &  \dfrac{(p_1\cdot p_2)^2 \big[q^2p_2^2 \ + \ 2 \, (p_2\cdot q)^2\big] 
	  \ + \ p_1^2p_2^2 (p_2\cdot q)^2 \ + \ 3\,p_2^4 (p_1\cdot q)^2  
	  \ - \ 6 \,p_2^2(p_1\cdot q)(p_2\cdot q)(p_1\cdot p_2) \ - \ 
	  q^2 p_1^2 p_2^4}{2\, \big[(p_1\cdot p_2)^2 \ - \ p_1^2p_2^2\big]^2} 
	\nonumber\\
   \beta_c & = & \dfrac{(p_1\cdot p_2)^2 \big[q^2p_1^2 \ + \ 2 \,(p_1\cdot q)^2\big] 
	  \ + \ p_1^2p_2^2 (p_1\cdot q)^2 \ + \ 3\,p_1^4 (p_2\cdot q)^2 
	  \ - \ 6 \,p_1^2 (p_1\cdot q)(p_2\cdot q)(p_1\cdot p_2) \ - \ q^2 p_1^4 p_2^2}
           {2\, \big[(p_1\cdot p_2)^2 \ - \ p_1^2p_2^2\big]^2} \nonumber\\
    \beta_d & = &    \Big\{ 4 \, (p_1\cdot q)(p_2\cdot q)(p_1\cdot p_2)^2 \ + \ 
		    p_1^2p_2^2 \big[2\, (p_1\cdot q)(p_2\cdot q) \ + \ q^2(p_1\cdot p_2) \big] \nonumber\\
		    & & \hspace{2cm} - \
		    3\, (p_1\cdot p_2)\big[p_1^2 (p_2\cdot q)^2 \ + \ p_2^2 (p_1\cdot q)^2\big] \ - \ q^2 (p_1\cdot p_2)^3 \Big\} 
		    \dfrac{1}
		    {2\, \big[(p_1\cdot p_2)^2 \ - \ p_1^2p_2^2\big]^2} .
\end{eqnarray}

\end{widetext}



\begin{thebibliography}{100}

\bibitem{Fukushima:2010bq}
  K.~Fukushima and T.~Hatsuda,
  Rept.\ Prog.\ Phys.\  {\bf 74} (2011) 014001
  doi:10.1088/0034-4885/74/1/014001
  [arXiv:1005.4814 [hep-ph]].

\bibitem{Villafane:2016ukb}
  M.~F.~Izzo Villafañe, D.~Gómez Dumm and N.~N.~Scoccola,
  Phys.\ Rev.\ D {\bf 94} (2016) no.5,  054003
  doi:10.1103/PhysRevD.94.054003
  [arXiv:1602.06984 [hep-ph]].
  
\bibitem{Carlomagno:2018tyk} 
  J.~P.~Carlomagno,
  Phys.\ Rev.\ D {\bf 97}, no. 9, 094012 (2018)
  doi:10.1103/PhysRevD.97.094012
  [arXiv:1803.03235 [hep-ph]].

\bibitem{Noguera:2008}
 S. Noguera, N. N. Scoccola,
 Phys.\ Rev.\  D {\bf 78}, 114002 (2008).

\bibitem{Carlomagno:2013ona} 
  J.~P.~Carlomagno, D.~G\'omez Dumm and N.~N.~Scoccola,
  Phys.\ Rev.\ D {\bf 88}, no. 7, 074034 (2013)
  doi:10.1103/PhysRevD.88.074034
  [arXiv:1305.2969 [hep-ph]].

\bibitem{GomezDumm:2001fz}
  D.~G\'omez Dumm and N.~N.~Scoccola,
  Phys.\ Rev.\ D {\bf 65}, 074021 (2002);
  Phys.\  Rev.\ C {\bf 72}, 014909 (2005). 

\bibitem{Hell:2011ic} 
  T.~Hell, K.~Kashiwa and W.~Weise,
  Phys.\ Rev.\ D {\bf 83}, 114008 (2011)
  doi:10.1103/PhysRevD.83.114008
  [arXiv:1104.0572 [hep-ph]].
  
\bibitem{Contrera:2010kz}
  G.~A.~Contrera, M.~Orsaria and N.~N.~Scoccola,
  Phys.\ Rev.\ D {\bf 82} (2010) 054026
  doi:10.1103/PhysRevD.82.054026
  [arXiv:1006.4639 [hep-ph]].
  
\bibitem{Hell:2008cc} 
  T.~Hell, S.~Roessner, M.~Cristoforetti and W.~Weise,
  Phys.\ Rev.\ D {\bf 79}, 014022 (2009)
  doi:10.1103/PhysRevD.79.014022
  [arXiv:0810.1099 [hep-ph]].
  
\bibitem{Radzhabov:2010dd} 
  A.~E.~Radzhabov, D.~Blaschke, M.~Buballa and M.~K.~Volkov,
  Phys.\ Rev.\ D {\bf 83}, 116004 (2011)
  doi:10.1103/PhysRevD.83.116004
  [arXiv:1012.0664 [hep-ph]].

\bibitem{Ripka:1997zb} 
  G.~Ripka,
  Oxford, UK: Clarendon Pr. (1997) 205 p 

\bibitem{GomezDumm:2004sr} 
  D.~Gomez Dumm and N.~N.~Scoccola,
  Phys.\ Rev.\ C {\bf 72}, 014909 (2005)
  doi:10.1103/PhysRevC.72.014909
  [hep-ph/0410262].
  
\bibitem{Bratovic:2012qs} 
  N.~M.~Bratovic, T.~Hatsuda and W.~Weise,
  Phys.\ Lett.\ B {\bf 719}, 131 (2013)
  doi:10.1016/j.physletb.2013.01.003
  [arXiv:1204.3788 [hep-ph]].

\bibitem{Contrera:2012wj} 
  G.~A.~Contrera, A.~G.~Grunfeld and D.~B.~Blaschke,
  Phys.\ Part.\ Nucl.\ Lett.\  {\bf 11}, 342 (2014)
  doi:10.1134/S1547477114040128
  [arXiv:1207.4890 [hep-ph]].

\bibitem{tHooft:1977nqb}
  G.~'t Hooft,
  Nucl.\ Phys.\ B {\bf 138} (1978) 1.
  doi:10.1016/0550-3213(78)90153-0
  
\bibitem{Polyakov:1978vu}
  A.~M.~Polyakov,
  Phys.\ Lett.\  {\bf 72B} (1978) 477.
  doi:10.1016/0370-2693(78)90737-2
  
\bibitem{Ratti:2005jh}
  C.~Ratti, M.~A.~Thaler and W.~Weise,
  Phys.\ Rev.\  D {\bf 73}, 014019 (2006).

\bibitem{Scavenius:2002ru}
  O.~Scavenius, A.~Dumitru and J.~T.~Lenaghan,
  Phys.\ Rev.\ C {\bf 66}, 034903 (2002).

\bibitem{Schaefer:2007pw} 
  B.~J.~Schaefer, J.~M.~Pawlowski and J.~Wambach,
  Phys.\ Rev.\ D {\bf 76}, 074023 (2007)
  doi:10.1103/PhysRevD.76.074023
  [arXiv:0704.3234 [hep-ph]].

\bibitem{Parappilly:2005ei}
  M.~B.~Parappilly, P.~O.~Bowman, U.~M.~Heller, D.~B.~Leinweber, A.~G.~Williams and J.~B.~Zhang,
  Phys.\ Rev.\ D {\bf 73} (2006) 054504
  doi:10.1103/PhysRevD.73.054504
  [hep-lat/0511007].
  
\bibitem{Ebert:1985kz} 
  D.~Ebert and H.~Reinhardt,
  Nucl.\ Phys.\ B {\bf 271}, 188 (1986).
  doi:10.1016/0550-3213(86)90359-7, 10.1016/S0550-3213(86)80009-8
    
\bibitem{Bernard:1993rz} 
  V.~Bernard, U.~G.~Meissner and A.~A.~Osipov,
  Phys.\ Lett.\ B {\bf 324}, 201 (1994)
  doi:10.1016/0370-2693(94)90408-1
  [hep-ph/9312203].
  
\bibitem{Contrera:2009hk}
  G.~A.~Contrera, D.~G\'omez Dumm and N.~N.~Scoccola,
  Phys.\ Rev.\ D {\bf 81}, 054005 (2010).
  
\bibitem{Bowler:1994ir} 
  R.~D.~Bowler and M.~C.~Birse,
  Nucl.\ Phys.\ A {\bf 582}, 655 (1995)
  doi:10.1016/0375-9474(94)00481-2
  [hep-ph/9407336].
  
\bibitem{GomezDumm:2006vz} 
  D.~Gomez Dumm, A.~G.~Grunfeld and N.~N.~Scoccola,
  Phys.\ Rev.\ D {\bf 74}, 054026 (2006)
  doi:10.1103/PhysRevD.74.054026
  [hep-ph/0607023].

\bibitem{Noguera:2008cm} 
  S.~Noguera and N.~N.~Scoccola,
  Phys.\ Rev.\ D {\bf 78}, 114002 (2008)
  doi:10.1103/PhysRevD.78.114002
  [arXiv:0806.0818 [hep-ph]].

\bibitem{Weldon:1991ei} 
  H.~A.~Weldon,
  Annals Phys.\  {\bf 214}, 152 (1992).
  doi:10.1016/0003-4916(92)90065-T
    
\bibitem{Tanabashi:2018oca} 
  M.~Tanabashi {\it et al.} [Particle Data Group],
  Phys.\ Rev.\ D {\bf 98}, no. 3, 030001 (2018).
  doi:10.1103/PhysRevD.98.030001

\bibitem{Bazavov:2016uvm} 
  A.~Bazavov, N.~Brambilla, H.-T.~Ding, P.~Petreczky, H.-P.~Schadler, A.~Vairo and J.~H.~Weber,
  Phys.\ Rev.\ D {\bf 93}, no. 11, 114502 (2016)
  doi:10.1103/PhysRevD.93.114502
  [arXiv:1603.06637 [hep-lat]].
    
\bibitem{GomezDumm:2005hy} 
  D.~Gomez Dumm, D.~B.~Blaschke, A.~G.~Grunfeld and N.~N.~Scoccola,
  Phys.\ Rev.\ D {\bf 73}, 114019 (2006)
  doi:10.1103/PhysRevD.73.114019
  [hep-ph/0512218].

\bibitem{Contrera:2007wu} 
  G.~A.~Contrera, D.~Gomez Dumm and N.~N.~Scoccola,
  Phys.\ Lett.\ B {\bf 661}, 113 (2008)
  doi:10.1016/j.physletb.2008.01.069
  [arXiv:0711.0139 [hep-ph]].

\bibitem{Carlomagno:2015nsa} 
  J.~P.~Carlomagno, D.~Gomez Dumm and N.~N.~Scoccola,
  Phys.\ Rev.\ D {\bf 92}, no. 5, 056007 (2015)
  doi:10.1103/PhysRevD.92.056007
  [arXiv:1507.01560 [hep-ph]].

\bibitem{IzzoVillafane:2016jnx} 
  M.~F.~Izzo Villafañe and D.~Gomez Dumm,
  J.\ Phys.\ Conf.\ Ser.\  {\bf 706}, no. 4, 042013 (2016).
  doi:10.1088/1742-6596/706/4/042013

\bibitem{Blaschke:1984yj}
  D.~Blaschke, F.~Reinholz, G.~Ropke and D.~Kremp,
  Phys.\ Lett.\  {\bf 151B} (1985) 439.
  doi:10.1016/0370-2693(85)91673-9
  
\bibitem{Hufner:1996pq}
  J.~Hufner, S.~P.~Klevansky and P.~Rehberg,
  Nucl.\ Phys.\ A {\bf 606} (1996) 260.
  doi:10.1016/0375-9474(96)00206-0
  
\bibitem{Ayala:2012ch} 
  A.~Ayala, C.~A.~Dominguez, M.~Loewe and Y.~Zhang,
  Phys.\ Rev.\ D {\bf 86}, 114036 (2012)
  doi:10.1103/PhysRevD.86.114036
  [arXiv:1210.2588 [hep-ph]].

\bibitem{McLerran:2007qj} 
  L.~McLerran and R.~D.~Pisarski,
  Nucl.\ Phys.\ A {\bf 796}, 83 (2007)
  doi:10.1016/j.nuclphysa.2007.08.013
  [arXiv:0706.2191 [hep-ph]].
  
\bibitem{McLerran:2008ua} 
  L.~McLerran, K.~Redlich and C.~Sasaki,
  Nucl.\ Phys.\ A {\bf 824}, 86 (2009)
  doi:10.1016/j.nuclphysa.2009.04.001
  [arXiv:0812.3585 [hep-ph]].

\bibitem{Abuki:2008nm} 
  H.~Abuki, R.~Anglani, R.~Gatto, G.~Nardulli and M.~Ruggieri,
  Phys.\ Rev.\ D {\bf 78}, 034034 (2008)
  doi:10.1103/PhysRevD.78.034034
  [arXiv:0805.1509 [hep-ph]].

\end{thebibliography}
\end{document}